\documentclass[groupaddress,twocolumn,secnumarabic,
amssymb,amsmath,nobibnotes,aps,siunitx,prd,showkeys,showpacs,nofootinbib]{revtex4-1}%
\usepackage{graphicx}
\usepackage{epsf}
\usepackage{slashed}
\providecommand{\U}[1]{\protect\rule{.1in}{.1in}}

\newcommand{\be}{\begin{equation}}
\newcommand{\ee}{\end{equation}}
\newcommand{\CC}{\Lambda}

\newcommand{\mincir}{\raise
-3.truept\hbox{\rlap{\hbox{$\sim$}}\raise4.truept\hbox{$<$}\ }}
\newcommand{\magcir}{\raise
-3.truept\hbox{\rlap{\hbox{$\sim$}}\raise4.truept\hbox{$>$}\ }}

\newcommand{\rL}{\rho_{\Lambda}}

\newcommand{\vp}{\varphi}
\newcommand{\ka}{\kappa}

\begin{document}


\title{Starobinsky-like inflation and running vacuum in the context of Supergravity}


\author{Spyros Basilakos$^a$, Nick E. Mavromatos$^{b,c}$ and Joan Sol\`a$^d$}

\affiliation{$^a$Academy of Athens, Research Center for Astronomy and Applied Mathematics, Soranou Efessiou 4, 115 27 Athens, Greece. \\
$^b$Theoretical Particle Physics and Cosmology Group, Physics Department, King's College London, Strand, London WC2R 2LS. \\
 $^c$ Theoretical Physics Department, CERN, Geneva CH-1211 Geneva 23, Switzerland\\
 $^{d}$Departament de F\'\i sica Qu\`antica i Astrof\'\i sica, and Institute of Cosmos Sciences (ICCUB), Univ. de Barcelona, Av. Diagonal 647 E-08028 Barcelona, Catalonia, Spain.}


\begin{abstract}
\vspace{0.2cm} We describe the primeval inflationary phase of the
early Universe within a quantum field theoretical (QFT) framework
that can be viewed as the effective action of vacuum decay in the
early times. Interestingly enough, the model accounts for the
``graceful exit'' of the inflationary phase into the standard
radiation regime.  The underlying QFT framework considered here is Supergravity (SUGRA),
more specifically an existing formulation in which the
Starobinsky-type inflation (de-Sitter background) emerges from the
quantum corrections to the effective action after integrating out
the gravitino fields in their (dynamically induced) massive phase.
We also demonstrate that the structure of the effective action in
this model is consistent with the generic idea of renormalization group
(RG) running of the cosmological parameters, specifically it follows from the corresponding RG equation for the vacuum energy density as a
function of the Hubble rate, $\rho_{\Lambda}(H)$. Overall our
combined approach amounts to a concrete-model realization of
inflation triggered by vacuum decay in a fundamental physics
context which, as it turns out, can also be extended for the remaining epochs of the cosmological evolution until the current dark energy era.

\end{abstract}

\pacs{98.80.-k, 95.35.+d, 95.36.+x} \keywords{Cosmology; inflation;
vacuum}\maketitle

\hyphenation{tho-rou-ghly in-te-gra-ting e-vol-ving con-si-de-ring
ta-king me-tho-do-lo-gy fi-gu-re}

\section{Introduction \label{sec:1}}
In the last two years we have witnessed extraordinary developments
on experimental tests of inflationary models~\cite{encyclo}, based
on studies of photons in  the cosmic microwave background radiation.
In particular, the results of Planck collaboration~\cite{Planck} and
the associated non-observation of B-mode polarizations of primordial
light fluctuations, have imposed very stringent restrictions on
single scalar-field models of slow-roll inflation, allowing
basically models with very low tensor-to-scalar fluctuation ratio $
r = n_T/n_s \ll 1$, with a scalar spectral index $n_s \simeq 0.96 $
and no appreciable running. In fact the upper bound set by Planck
Collaboration~\cite{Planck} on this ratio, as a consequence of the
non-observation of B-modes, is $r < 0.11$, but their favored regions
point towards $r \le 10^{-3}$.  This is a
feature that characterizes the so-called Starobinsky-type (or
$R^2$-inflation, with $R$ denoting the scalar space-time curvature)
inflationary models~\cite{staro}. The estimated energy scale $E_I$
of inflation, which in inflaton-type models is related to the --
approximately constant-- scalar potential during inflation through
$E_I= V_I^{1/4}$,  reads~\cite{encyclo}:
\begin{equation}\label{hubblebicep2}
E_I \; = \; \Big(3\, H^2_I M_{\rm Pl}^2 \Big)^{1/4} \simeq 2.1
\times 10^{16} \times \Big(\frac{r}{0.20}\Big)^{1/4} \, {\rm GeV}~,
\end{equation}
where $M_{\rm Pl} =1/\sqrt{8\pi G}\simeq 2.43 \times 10^{18}~{\rm GeV}$
is the reduced Planck mass ($G$ being the Newtonian constant). The
upper bound $r < 0.11$ placed by the Planck Collaboration implies
\begin{equation}\label{upperH}
H_I  = 1.05 \, \Big(\frac{r}{0.20} \Big)^{1/2}\times 10^{14}~{\rm
GeV} \, \leq    0.78 \times 10^{14}~{\rm GeV}~.
\end{equation}
The above can be rephrased as $H_I/m_{P}   \leqslant
6.39\times10^{-6}$, where $m_P=1/G^{1/2} = \sqrt{8\pi}M_{\rm Pl} \simeq 1.22\times 10^{19}$
GeV is the Planck mass in natural units. This result is consistent
with the well known CMB bound $H_I/m_{P}\lesssim10^{-5}$ on the
temperature fluctuations induced by the tensor modes. As we will
see, the actual value of $H$ during inflation for the class of
models under study satisfies $H\lesssim H_I$, and hence the CMB
bound is preserved by them.

The recent joint BICEP2-Planck
analysis\,\cite{Bicep2-Planck2015} confirmed the early Planck result, namely
the likelihood curve for $r$ yields an upper limit $r <0.12$ at
$95\%$. Moreover, the present BICEP2-Planck  data are consistent with a scalar
spectral index $n_s \simeq 0.96 $ and no appreciable running, in
agreement with the previous Planck data~\cite{Planck}. Using the
aforementioned new upper limit $r_{\rm max}=0.12$, the Hubble
parameter during slow-roll inflation $H_I$ is estimated to be below
\begin{equation}\label{upperHbic}
H_I^{\rm Bicep2+Planck} \leq 0.81 \times 10^{14}~{\rm GeV}\,,
\end{equation}
and hence $H_I/m_{P}\leqslant 6.64\times10^{-6}$. Because of the low
significance of the new limit on $r$, the possibility that $r$ is
actually much smaller than the current upper limit  $r_{\rm max}$
remains as natural as it was before. In fact, nothing actually
prevents at present that the typical value of the tensor to scalar
ratio can be, for example, $r={\cal O}(10^{-3})$, and in this sense
the Starobinsky-type scenarios can still be considered as a serious
possibility to describe the inflationary universe.
Following this point of view, we continue in this paper with the
investigation of Starobinsky-like models as potential candidates for
realistic implementation of inflation compatible with the data.

In previous publications one of us (N.E.M.) with
collaborators~\cite{ahm} discussed the dynamical breaking of
supergravity (SUGRA) theories via gravitino condensation, and
demonstrated~\cite{ahmstaro} the compatibility of this scenario with
Starobinsky-like~\cite{staro} inflationary scenarios.
As we discussed, this phase is characterized  by the dynamical
emergence of a de-Sitter background. As argued in \cite{ahm}, the
Starobinsky-type inflation appears much more natural (from the point
of view of the order of the parameters involved) than a hill-top
inflation scenario~\cite{emdyno} in which the gravitino condensate
itself is the inflaton field. In the latter, very large values of
the wave function renormalization of the condensate field are
required to ensure slow-roll inflation if one insists on
(phenomenologically realistic) sub-Planckian supersymmetry breaking
scales. It is important to notice at this point that in the original
Starobinsky model~\cite{staro} the  $R^2$ terms crucial for
inflation arise from  the \emph{conformal anomaly} in the path
integral of massless (conformal) matter in a de Sitter background,
and thus their coefficient is arbitrary, and can only be fixed
phenomenologically. A similar, although not identical, situation
occurs in the context of \emph{anomaly-induced
inflation}\,\cite{SSinflation02,Fossil07}, where the term $R^2$ is
absent at the classical level but is generated from the conformal
anomaly. In this case, however, the coefficient of $\Box R$
(entering the $\beta$-functions and controlling the stability of
inflation) presents also some arbitrariness which can only be fixed
by a special renormalization condition. Par contrast, in the
considered SUGRA scenario, such terms arise in the one-loop
effective action of the gravitino condensate field, evaluated in a
de Sitter background, after integrating out massive gravitino
fields, whose mass was generated dynamically. The order of the de
Sitter cosmological constant, $\Lambda > 0$ that breaks
supersymmetry, and the gravitino mass are all evaluated dynamically
(self-consistently) in our approach from the minimization of the
effective potential. Thus, the resulting $R^2$ coefficient, which
determines the phenomenology of the inflationary phase,  is
calculable~\cite{ahmstaro}.

Also very important for our considerations is the framework of the
\textit{running vacuum model} (RVM)
~\cite{ShapSol,ShapSol09,SolStef05,Fossil07} -- see
\cite{JSPRev2013,SolGo2015a} and references therein for a recent
detailed exposition. The implications of these dynamical vacuum
models have recently been analyzed both for the early
universe\,\cite{LBS2014,LBS2013,BLS2013,Perico2013,SolGo2015a,SolaGRF2015} as well as
for the phenomenology of the current
universe\,\cite{GoSolBas2015,GoSol2015,Elahe2015} -- see also
\cite{BPS09,GrandeET11,FritzschSola2012,BasPolSol2012,BasSol2013,Cristina}
for previous analyses.

\phantom{} In regard to the early universe we emphasize
that the RVM defines a class of non-singular inflationary scenarios
with  graceful exit into the standard radiation regime. These models are related to Starobinsky inflation models, although
they are not equivalent. We will discuss in this paper the correspondence between them, and most particularly
with the dynamically broken SUGRA model with gravitino condensation that we have mentioned above.
It is especially remarkable that such specific implementation of
the SUGRA model leads, as we will show in this paper, to the
effective behavior of the RVM with calculable coefficients. In this way the former automatically
benefits from the successful consequences of the latter. Let us
mention that the RVM also provides some important clues for
alleviating the cosmological constant problem\,\cite{JSPRev2013}.

\phantom{} Finally, we should like to mention that the RVM's have been tested
against the wealth of accurate SNIa+BAO+$H(z)$+LSS+BBN+CMB data -- see \cite{JSPMG14} for a
recent summary review -- and they turn out to provide a quality fit that is significantly
better than the $\CC$CDM. This fact has become especially prominent in the light of the most
recent works\,\cite{SolGoJa2015,SolGoJa2016}. \phantom{} Therefore, there is every motivation
for further investigating these dynamical vacuum models from different perspectives, with the
hope of finding possible connections with fundamental aspects of the cosmic evolution.
In point of fact, this is the main aim of this work.

The structure of the article is as follows. The general framework of
the RVM is introduced in \ref{sec:running}. The basic theoretical
elements of the Starobinsky inflation are presented in section
\ref{sec:2}. The main properties of the dynamical breaking of local
SUGRA theory and its connection to Starobinsky-type inflation are
reviewed in section \ref{sec:3}. In section \ref{sec:4}  we
demonstrate how the RVM describes the effective framework of the
Starobinsky \cite{staro} and the dynamically broken SUGRA
\cite{emdyno} models at the inflationary epoch. Finally, our
conclusions are summarized in section \ref{sec:5}.

\section{Running vacuum: a natural arena for vacuum decay in cosmology} \label{sec:running}

It is the purpose of this work to go one step further from
demonstrating compatibility of the dynamically broken SUGRA
scenario\,\cite{ahm,ahmstaro} with inflation and discuss the
possibility of a dynamical evolution
of the inflationary
phase ground state to the standard radiation regime within the
context of the running vacuum model (RVM) of the cosmic evolution
utilizing an effective ``renormalization group (RG) approach'',
see\,\cite{ShapSol,Fossil07} -- and \cite{JSPRev2013,SolGo2015a} for
comprehensive expositions. Specifically, we wish to show that the
behavior of the aforementioned SUGRA scenario effectively mimics the
RVM. Once this link is elucidated, the general ``decaying'' vacuum
description inherent to the RVM formulation allows to smoothly
connect inflation to the standard
Fridman-Lema\^{\i}tre-Robertson-Walker (FLRW) radiation era,  which
subsequently proceeds into a matter and dark energy domination in
the present era, in which it still carries a mild dynamical behavior compatible with the current cosmological data\,\cite{GoSolBas2015,GoSol2015}. Such an expansion history of the
Universe has been put forward in previous works by the authors in
various collaborations and contexts, \emph{e.g}. non-equilibrium
string-inspired cosmologies~\cite{string} or conventional
field-theoretic cosmologies in which the above mentioned RVM is
extensively applied for the study of the early cosmic
history\,\cite{LBS2013,BLS2013,Perico2013,LBS2014}.

In the effective RG approach underlying the RVM one can write down
an evolution equation for the effective vacuum energy density
$\rL(t)=\rL(\mu_c(t))$, treated as a dynamical quantity whose cosmic
time evolution is inherited from its dependence on a characteristic
cosmic scale variable $\mu_c=\mu_c(t)$. This variable plays the role
of running (mass) scale of the renormalization group approach, and a
natural candidate for such scale in FLRW cosmology is the Hubble
parameter $H(t)$. Therefore the proposed RG equation is
~\cite{JSPRev2013}:
\begin{equation}\label{runningrho}
\frac{d\, \rho_\Lambda (t)}{d\, {\rm ln}H^2 } = \frac{1}{(4\pi)^2}
\sum_i \Big[a_i M_i^2 H^2 + b_i H^4 + c_i \frac{H^6}{M_i^2} + \dots
\Big]
\end{equation}
In general $\mu_c^2$ can be associated to a linear combination of
$H^2$  and $\dot{H}$ and the variety of terms appearing on the
r.h.s. of (\ref{runningrho}) can be richer\,\cite{SolGo2015a}, but
the canonical possibility is the previous one and hereafter we
restrict to it. The coefficients $a_i, b_i, c_i \dots$ appearing in
(\ref{runningrho}) are dimensionless and receive contributions from
loop corrections of boson and fermion matter fields with different
masses $M_i$. It must be stressed that the general covariance of the
action\,\cite{Fossil07,ShapSol09,SolStef05} necessitates the appearance of
\emph{only even} powers of the (cosmic-time $t$ dependent) Hubble
parameter $H(t)$ on the right-hand-side of (\ref{runningrho}). For a
specific framework where the above  RG is concretely realized and
the $\beta$-function coefficients can be computed, see
\cite{Fossil07}.

We  note at this stage that, if the evolution of the Universe is
restricted to eras below the Grand Unified Theory (GUT) scale, then
for all practical purposes it is at most  the $H^4$ terms (those with dimensionless coefficients $b_i$) that can
contribute significantly. The $H^2$ term is of course negligible at
this point, and the higher powers of $H^n$ for $n=6,8,..$ are
suppressed by the corresponding inverse powers of the heavy masses $M_i$, which go to the denominator, as required by the decoupling theorem. In the
scenarios of dynamical breaking of local supergravity discussed in
ref.~\cite{ahm,ahmstaro,emdyno}, the breaking and the associated
inflationary scenarios could occur around the GUT scale, in
agreement with the inflationary phenomenology suggested by the Planck
satellite data~\cite{Planck}, provided Jordan-frame supergravity models
(with broken conformal symmetry) are used, in which the conformal
frame function
acquired, via appropriate dynamics, some non
trivial vacuum expectation value. For these situations, therefore,
corrections in (\ref{runningrho}) involving higher powers than $H^4$
will be ignored.

In the next sections, after revising the general framework of
Starobinsky inflation, we shall compute $\rho_\Lambda$ in such
supergravity models and study their evolution from the exit from the
Starobinsky inflationary phase, that occurs in the massive gravitino
phase, until today. The computation of $\rho_\Lambda$ will be made
via the corresponding calculation of the one-loop effective action
after massive gravitinos are integrated out in a path integral.
Then, an identification of the effective equation of state can be
derived by integrating (\ref{runningrho}), following the approach of
\,\cite{LBS2013,BLS2013,Perico2013,LBS2014}.
Before doing so, it is instructive to review first the emergence of
Starobinsky-type inflation.

\section{Generic Starobinsky Inflation \label{sec:2}}
Starobinsky inflation is the oldest model of inflation~\cite{staro},
prior to the traditional, scalar-field-based,  inflaton models. It
is characterized for being able to realize the de Sitter
(inflationary) phase from the gravitational field equations derived
from a four-dimensional action that includes higher curvature terms,
specifically of the type involving the quadratic curvature
correction $\sim R^2$ ~\cite{staro}~\footnote{Our metric signature is $(-, +, +, +)$ and the definitions of the Ricci
and Riemann curvature tensors are  $R_{\mu\nu} = R^\lambda_{\mu \lambda \nu}$  and  $R^\lambda_{\mu\nu\rho} = \partial_\nu \, \Gamma_{\mu\rho}^\lambda - \dots $, respectively, i.e. we follow the exact three-sign conventions $(+,+,+)$ of Misner-Thorn-Wheeler\,\cite{MTW73}.}:
\begin{eqnarray}\label{staroaction}
{\mathcal S} = \frac{1}{2 \, \kappa^2 } \, \int d^4 x \sqrt{-g}\,
\left(R  + \beta  \, R^2 \right) ~,~ \beta \equiv \frac{8\, \pi}{3\,
{\mathcal M}^2 }~.
\end{eqnarray}
Here $\kappa^2=8\pi G = 1/M_{\rm Pl}^2$ (in units of $\hbar=c=1$ we
are working on), ${\rm G}=1/m_P^2$ is Newton's (gravitational)
constant in four space-time dimensions, with $m_P$ the Planck mass,
and ${\mathcal M}$ is a constant of mass dimension one,
characteristic of the model. Notice that the curvature terms in the
action are just the dimension-4 combination $m_P^2R/16\pi+R^2
m_P^2/(6{\cal M}^2)$. With this normalization, ${\cal M}$ gives the
value of the so-called scalaron mass. The smaller is ${\cal M}$ in
Planck mass units (i.e. the larger is the dimensionless parameter
$m_P^2/{\cal M}^2$ in front of $R^2$) the longer is the inflationary
time (cf. Fig. 2 of \,\cite{SolGo2015a}). Of course ${\cal M}$ cannot be much below the natural scale of
inflation, and in fact it should be of the same order, i.e. ${\cal
M}\sim M_X$, where $M_X$ is some GUT scale below the Planck mass.
Typically $M_X\sim 10^{16}\,{\rm GeV}\sim 10^{-3} m_P$.

The most relevant feature of this model is that inflationary
dynamics is driven by the purely gravitational sector, through the
$R^2$ terms.
From a microscopic point of view, these terms can be  viewed as the
result of \emph{quantum fluctuations} (at one-loop level)  of
conformal (massless or high energy) \emph{matter fields} of various
spins, which have been integrated out in the relevant path integral
in a curved background space-time~\cite{loop}. The model in fact is
to be understood in the context of QFT in curved spacetime. The
quantum mechanics of this model, by means of \emph{tunneling} of the
Universe from a state of ``nothing'' to the inflationary phase of
ref.~\cite{staro} has been discussed in detail in \cite{vilenkin}.
The above considerations necessitate truncation to one-loop quantum
order and to curvature-square (four-derivative) terms, which implies
that there must be a region of validity for curvature invariants
such that $\mathcal{O}\big(R^2/m_P^4\big) \ll 1$. Recalling that $R
\sim 12 \, H_I^2$  in the inflationary phase (where $H_I$ is the
nearly constant Hubble rate in that phase), we observe that this is
indeed a condition satisfied in phenomenologically realistic
scenarios of inflation~\cite{encyclo,Planck}, for which the
inflationary Hubble scale $H_I$ is typically constrained to obey
(\ref{upperH}) (Planck data~\cite{Planck}) or (\ref{upperHbic})
(BICEP2 data~\cite{Bicep2}), which are at present essentially the
same.

\begin{figure}
\includegraphics[width=0.45\textwidth]{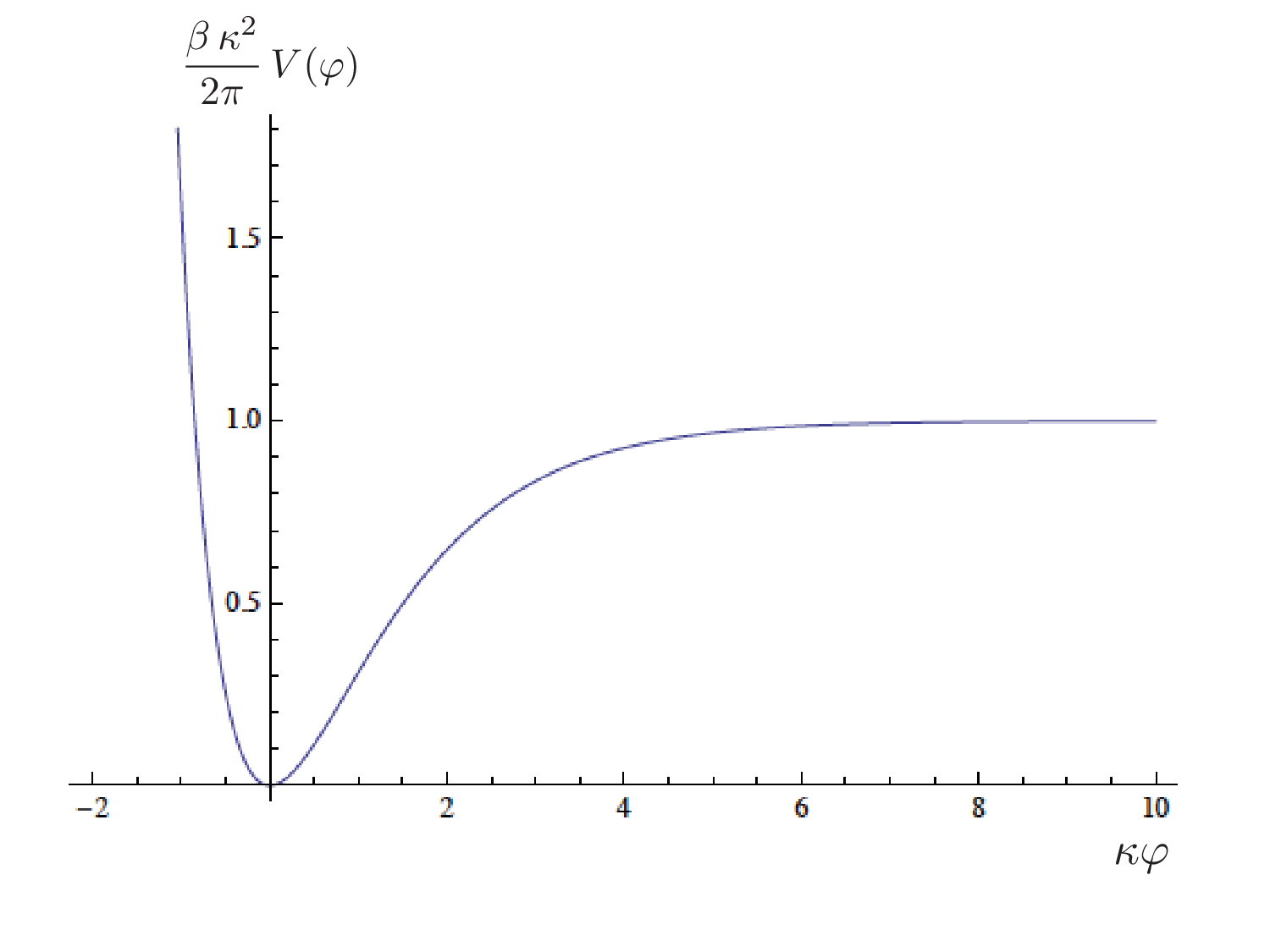}
\caption{The effective potential (\ref{staropotent}) of the
collective scalar field $\varphi$ that describes the one-loop
quantum fluctuations of matter fields, leading to the higher-order
scalar curvature corrections in the Starobinsky model for inflation
(\ref{staroaction}). Notice that according to
(\ref{staropotent}) we have $\beta=8\pi/3{\cal M}^{2}$.
The potential is sufficiently flat for
$\kappa\vp\gg1$ to ensure slow-roll conditions for inflation are
satisfied, in agreement with the Planck data, for appropriate values
of the scale $1/\sqrt{\beta} \propto {\mathcal M}$ (which sets the
overall scale of inflation in the model).}
\label{fig:potstar}
\end{figure}

Although the inflation in this model is not driven by fundamental
rolling scalar fields, nevertheless the model (\ref{staroaction})
(and for that matter, any other model where the Einstein-Hilbert
space-time Lagrangian density is replaced by an arbitrary function
$f(R)$ of the scalar curvature)  is \emph{conformally equivalent} to
that of an ordinary Einstein-gravity coupled to a scalar field with
a potential that drives inflation~\cite{whitt}. To see this, one
firstly linearises the $R^2$ terms in (\ref{staroaction}) by means
of an auxiliary (Lagrange-multiplier) field $\tilde \alpha (x)$,
before rescaling the metric by a conformal transformation and
redefining the scalar field (so that the final theory acquires
canonically-normalised Einstein and scalar-field terms):
\begin{eqnarray}\label{confmetric}
&&g_{\mu\nu} \rightarrow g^E_{\mu\nu} = \left(1 + 2 \, \beta \, {\tilde \alpha (x)} \right) \, g_{\mu\nu} ~, \nonumber \\
&& \tilde \alpha \left(x\right) \to \kappa\varphi (x) \equiv
\sqrt{\frac{3}{2}} \, {\rm ln} \, \left(1 + 2\, \beta \, {\tilde
\alpha \left(x\right)} \right)~,
\end{eqnarray}
where again $\kappa=\sqrt{8\pi G}$. These steps may be understood
schematically via
\begin{align}\label{steps}
    &\int d^4 x \sqrt{-g}\,  \left( R  + \beta  \, R^2 \right)  \\
    &\hookrightarrow\int d^4 x \sqrt{-g}\,  \left(  \left(1 + 2\, \beta \, \tilde \alpha \left(x\right) \right) \, R  -  \beta  \, {\tilde \alpha (x)}^2 \right)\nonumber  \\
    &\hookrightarrow\int d^4 x \sqrt{-g^E}\,  \left(R^E -  g^{E\, \mu\, \nu} \, \partial_\mu \, \varphi \, \partial_\nu \, \varphi -
V(\varphi ) \right)~,\nonumber
\end{align}
where the arrows have the meaning that the corresponding actions
appear in the appropriate path integrals. The ensuing effective
potential $V_{\rm eff}(\varphi)$ is given by:
\begin{eqnarray}\label{staropotent}
 V_{\rm eff}(\varphi ) = \frac{3{\cal M}^2\left( 1 - e^{-\sqrt{\frac{2}{3}} \, \ka\varphi } \right)^2}{4\,\ka^2} \,  ~.
\end{eqnarray}
One can check that the mass of the scalaron, which can be seenn as
the new gravitational
degree of freedom that the conformal transformation was able to
elucidate from the Starobinsky action, is indeed given by the
parameter ${\cal M}=\sqrt{8\pi/3\beta}$:
\begin{equation}\label{eq:scalaronMass}
\left.\frac{d^2 V_{\rm eff}(\vp)}{d\vp^2}\right|_{\vp=0}={\cal
M}^2\,.
\end{equation}
 Note that for $\vp=0$ one has $\tilde{\alpha}=0$ and the two conformally equivalent metrics coincide at this point. The effective potential for the scalar d.o.f. that conformally replaces the effect of the $R^2$ term is plotted in Fig.~\ref{fig:potstar}.
We observe that $V(\vp)$  is sufficiently flat for
$\ka\vp=\vp/M_{Pl}\gg 1$ (i.e. for sufficiently large values of
$\vp$ as compared to the reduced Planck scale) to produce
phenomenologically acceptable inflation. Obviously the scalaron
field $\varphi$ is effectively playing the role of the inflaton in
this context. The difference with the usual inflaton is that $\vp$
is not a new scalar d.o.f. imported from outside the gravitational
action, but just an integral part of it, namely, it is just a
gravitational d.o.f. that describes in an effective (and very
convenient) way the $\sim R^2$ term of (\ref{staroaction}). The
Starobinsky model based on the action (\ref{staroaction})  indeed
fits excellently the Planck data on inflation~\cite{Planck}, and
also the corresponding data from the joint BICEP2-Planck
analysis\,\cite{Bicep2-Planck2015}.

Quantum-gravity corrections in the original Starobinsky model
(\ref{staroaction}) have been considered recently in \cite{copeland}
from the point of view of an \emph{exact renormalisation-group}
analysis~\cite{litim}. It was shown that the \emph{non-perturbative}
beta-functions for the `running' of Newton's `constant' G and the
dimensionless inverse $R^2$ coupling  $\kappa^2\beta^{-1}\sim {\cal
M}^2/M_{Pl}^2$ in (\ref{staroaction}) imply an \emph{asymptotically
safe} Ultraviolet (UV) fixed point for the former (that is, G($k \to
\infty$) $ \to $ constant, for some 4-momentum cutoff scale $k$),
in the spirit of Weinberg~\cite{weinberg}, and an attractive
\emph{asymptotically-free} ($\kappa^2\beta^{-1} (k \to \infty) \to
0$) point for the latter. In this sense, the smallness of the (inverse) $R^2$
coupling, required for agreement with inflationary
observables~\cite{Planck}, is naturally ensured by the presence of
the asymptotically free UV fixed point.

The agreement of the model of \cite{staro} with the Planck data
triggered an enormous interest in the current literature, and indeed Starobinsky inflation has been
revisited from various points of view, such as its
connection with no-scale supergravity~\cite{olive} and
(super)conformal versions of supergravity and related
areas~\cite{sugrainfl}. In the latter works however the Starobinsky
scalaron field is fundamental, arising from the appropriate
scalar component of some chiral superfield that appears in the
superpotentials of the model.

Although of great value, illuminating a strong connection between
supergravity models and inflationary physics, and especially for
explaining the low-scale of inflation compared to the Planck scale,
these works contradict the original spirit of the Starobinsky model
(\ref{staroaction}) where, as mentioned previously, the higher
curvature corrections are viewed as arising from quantum
fluctuations of matter fields in a curved space-time background such
that inflation is driven by the pure gravity sector in the absence
of fundamental scalars. On the other hand, the scenario of
\cite{ahmstaro}, in which a Starobinsky-type inflation arises in the
massive gravitino phase of SUGRA models, after integrating out the
massive degrees of freedom is in the same spirit of Starobinsky, and
even better, in the sense that the model does not have to assume
dominance of conformal matter during inflation.

We next proceed to summarize the construction of the one-loop
effective action of the massless degrees of freedom after massive
gravitino integration in this dynamically-broken SUGRA model with
spontaneous breaking of global supersymmetry (SUSY)\,\cite{ahm}.

\section{Starobinsky-type inflation in Dynamically Broken SUGRA \label{sec:3}}

Dynamical breaking of SUGRA, in the sense of the generation of a
mass for the gravitino field $\psi_\mu$, whilst the gravitons remain
massless, occurs in the model as a result of the four-gravitino
interactions characterizing the SUGRA action, arising from the
torsionful contributions of the spin connection, characteristic of
local supersymmetric theories.

Our starting point is the $\mathcal{N}=1$ $D=4$ (on-shell) action for `minimal' Poincar\'e supergravity in the second order formalism~\cite{Freedman}:
 \begin{align}\label{sugraction}
	&S_{\rm{SG}}=\int d^4x \,e \left(\frac{1}{2\kappa^2}R\left(e\right)-\overline\psi_\mu\gamma^{\mu\nu\rho}D_\nu\psi_\rho+\mathcal{L}_{\rm torsion}\right),\\\nonumber
	&\kappa^2=8\pi G\,,
	 \quad\gamma^{\mu\nu\rho}=\frac{1}{2}\left\{\gamma^\mu,\gamma^{\nu\rho}\right\}\,,
	\quad \gamma^{\nu\rho}=\frac{1}{2}\left[\gamma^\nu,\gamma^\rho\right]\,,
\end{align}
where $R(e)$ and $D_\nu\psi_\rho\equiv\partial_\nu\psi_\rho+\frac{1}{4}\omega_{\nu ab}\left(e\right)\gamma^{ab}\psi_\rho$ are defined via the torsion-free connection; and, given the gauge condition $\gamma\cdot\psi=0$,
\begin{align}\label{torsion}
	\mathcal{L}_{\rm torsion}=-\frac{\kappa^2}{8}\, \left(\overline\psi^\rho\gamma^\mu\psi^\nu\right)\, \Big(\overline\psi_\rho\gamma_\mu\psi_\nu+2\overline\psi_\rho\gamma_\nu\psi_\mu\Big)~,
\end{align}
arising from the fermionic torsion parts of the spin connection.
Extending the action off-shell requires the addition of auxiliary fields to balance the graviton and gravitino degrees of freedom.
These fields however are non-propagating and may only contribute through the development of scalar vacuum expectation values, which would ultimately be resummed into the cosmological constant.

Making further use of the above gauge condition together with the Fierz identities (as detailed in \cite{ahm}), we may write
{\small \begin{align}\label{fierz}
		\frac{\mathcal{L}_{\rm torsion}}{\kappa^2} =  \lambda_{\rm S} \left(\overline\psi^\rho\psi_\rho\right)^2
		+\lambda_{\rm PS} \left(\overline\psi^\rho\gamma^5\psi_\rho\right)^2
		+\lambda_{\rm PV} \left(\overline\psi^\rho\gamma^5\gamma_\mu\psi_\rho\right)^2
\end{align}}where the couplings $\lambda_{\rm S}$, $\lambda_{\rm PS}$ and $\lambda_{\rm PV}$ express the freedom we have to rewrite each quadrilinear in terms of the others via Fierz transformation.
This freedom in turn leads to a known ambiguity in the context of (perturbative) mean field theory \cite{Wetterich} and can only be resolved by a non-perturbative treatment.

Specifically, we wish to linearise these four-fermion interactions via suitable auxiliary fields, e.g.
\begin{align}\label{aux}
	\frac{1}{4} \kappa ^2 \lambda_S \left(\overline\psi^\rho\psi_\rho\right)^2\sim \sigma\, \kappa \, \sqrt{\lambda_S}\, \left(\overline\psi^\rho\psi_\rho\right)-\sigma^2\,,
\end{align}	
where the equivalence (at the level of the action) follows as a consequence of the subsequent Euler-Lagrange equation for the auxiliary scalar $\sigma$.
Our task is then to look for a non-zero vacuum expectation value $\langle\sigma\rangle$ which would induce as an effective mass $m_{3/2}\sim \sigma\, \kappa \, \sqrt{\lambda_S}$ for the gravitino.
This is however complicated by the fact that our coupling $\lambda_{\rm S}$ into this particular channel is, by virtue of Fierz transformations, ambiguous at a perturbative level and, as mentioned, in order to fix them a fully non-perturbative treatment of SUGRA-like models would be required, which are not currently at hand. Nevertheless, there is another way out~\cite{emdyno,ahm} whereby the Fierz ambiguities may be absorbed by dilaton-expectation-value shifts in an extension of ${\mathcal N}=1$ SUGRA which incorporates local supersymmetry in the Jordan frame, enabled by an associated dilaton superfield~\cite{confsugra}.
The (logarithm of the) scalar component $\varphi$ of the latter can be either a fundamental space-time scalar mode of the gravitational multiplet, i.e. the trace of the graviton (as happens, for instance, in supergravity models that appear in the low-energy limit of string theories), or a composite scalar field constructed out of matter multiplets.
In the latter case these could include the standard model fields and their superpartners that characterise the Next-to-Minimal Supersymmetric Standard Model~\cite{nmssm}, which can be consistently incorporated in such Jordan frame extensions of SUGRA.

Upon appropriate breaking of conformal symmetry, induced by specific dilaton potentials (which we do not discuss here), one may then assume that the dilaton field acquires a non-trivial vacuum expectation value $\langle \phi \rangle \ne 0 $, thus absorbing any ambiguities in the value of the appropriate coefficient $\lambda_S$ induced by Fierz (\ref{fierz}).
One consequence of this is then that in the broken conformal symmetry phase, the resulting supergravity sector, upon passing (via appropriate field redefinitions) to the Einstein frame is described by an action of the form (\ref{sugraction}), but with the coupling of the gravitino four-fermion interaction terms being replaced by
\begin{equation}\label{tildcoupl}
\tilde \kappa^2 \equiv \lambda_S \, \kappa^2 =  e^{-4\langle\phi\rangle}\kappa^2~,
\end{equation}
while the Einstein term in the action carries the standard gravitational coupling $1/2\kappa^2$. For phenomenological reasons, associated with gravitino masses in the ball-park of Grand Unified Theory (GUT) scales, one must have $\tilde \kappa \gg \kappa$. This is assumed to be guaranteed by appropriate microscopic dilaton potentials that break the (super)conformal symmetry of the Jordan-frame SUGRA appropriately.

To induce the super-Higgs effect~\cite{DeserZumino} we couple to the action (\ref{sugraction}) the Goldstino associated to global supersymmetry breaking via the addition of
\begin{align}\label{goldstino}
	 \mathcal{L}_\lambda=f^2\det\left(\delta_{\mu\nu}+\frac{i}{2f^2}\overline\lambda\gamma_\mu\partial_\nu\lambda\right)\bigg|_{\gamma\cdot\psi=0}=f^2+\dots\,,
\end{align}	
where $\lambda$ is the Goldstino, $\sqrt{f}$ expresses the scale of global supersymmetry breaking, and \dots\, represents higher order terms which may be neglected in our weak-field expansion of the determinant.
It is worth emphasising at this point the universality of \eqref{goldstino}; any model containing a Goldstino may be related to $\mathcal{L}_\lambda$ via a non-linear transformation \cite{Komargodski}, and thus the generality of our approach is preserved.

Upon the aforementioned gauge choice for the gravitino field
$ \gamma^\mu \psi_\mu = 0~,$
and an appropriate redefinition, one may eliminate any presence of the Goldstino field from the final effective
action describing the dynamical breaking of local supersymmetry, except the cosmological constant term $f^2$ in (\ref{goldstino}), which serves as a reminder of the pertinent scale of supersymmetry breaking. The non-trivial energy scale this introduces, along with the disappearance (through field redefinitions) of the Goldstino field from the physical spectrum and the concomitant development of a gravitino mass, characterises the super-Higgs effect.

The linearisation of the four-gravitino terms (\ref{aux}), when combined with the $f^2$ term of the super-Higgs effect implies a tree-level
cosmological constant
\begin{align}\label{l0tree}
	 \frac{\Lambda_0}{\kappa^2} 	
	 \equiv \sigma ^2  -f^2 ~,
\end{align}
which \emph{must} be \emph{negative} due to the incompatibility of supergravity with de Sitter vacua (notice that in our conventions both $\sigma$ and $f$ have dimension $+2$ in natural
units).

\begin{figure}
\centering
\includegraphics[width=0.40\textwidth]{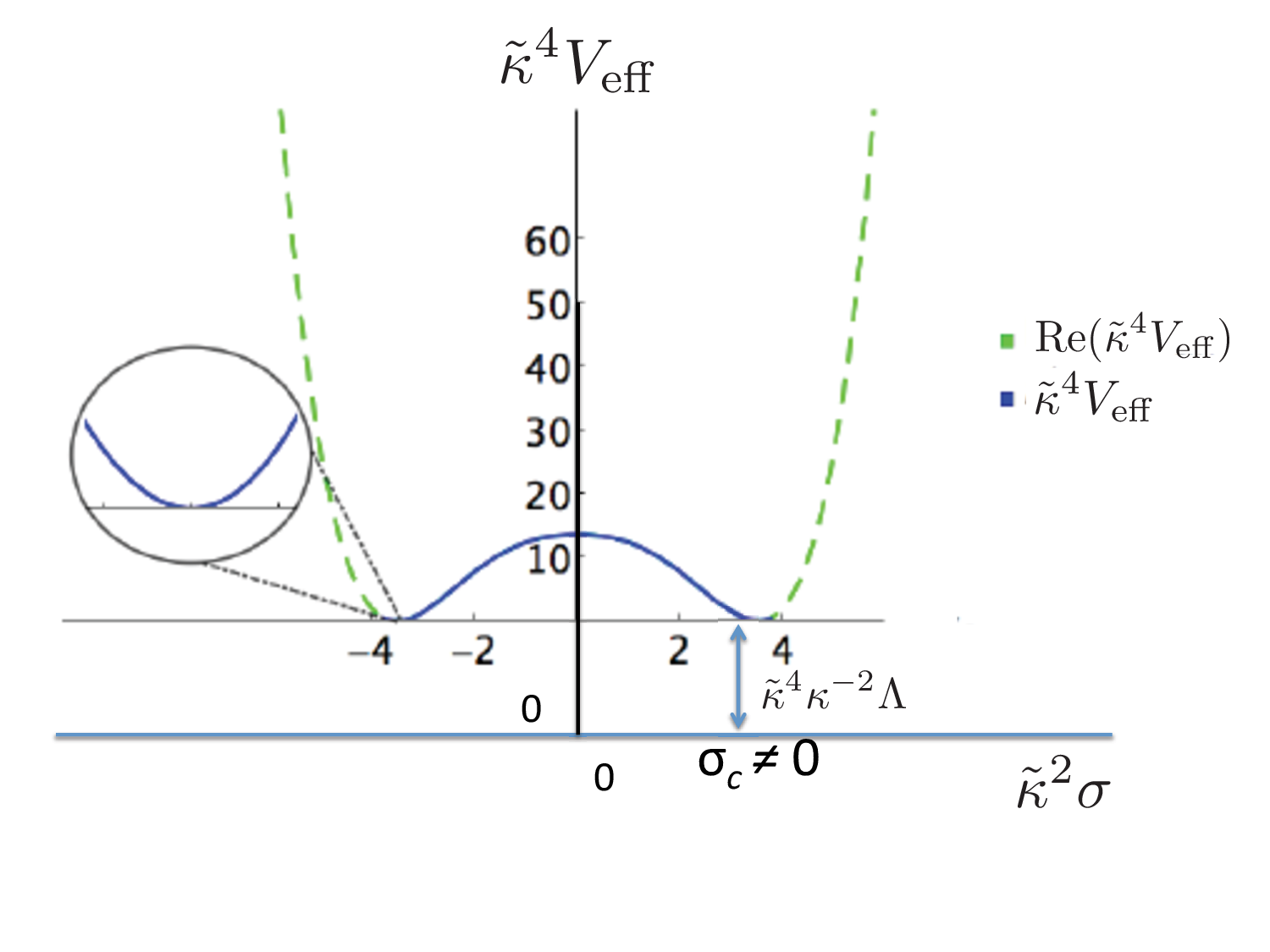}
\caption{Generic shape of the one-loop effective potential (in
dimensionless units ${\tilde \kappa}^4\,V_{\rm eff}$ , where
${\tilde \kappa}$ is the conformally rescaled gravitational
coupling, see the text) for the gravitino condensate field
$\sigma$ in dynamically broken (conformal) Supergravity models in
the presence of a non-trivial de Sitter background with cosmological
constant $\Lambda > 0$~\cite{ahmstaro}. The Starobinsky inflationary
phase is associated with fluctuations of the condensate and
gravitational field modes near the non-trivial minimum of the
potential, where the condensate $\sigma_c \ne 0$, and the potential
assumes the value $\Lambda > 0$, consistent with supersymmetry
breaking. The dashed green lines denote ``forbidden'' areas of the
condensate field values, violating the condition (\ref{absimparts}),
for which imaginary parts appear in the effective potential, thereby
destabilizing the broken symmetry phase.} \label{fig:pot2}
\end{figure}

The one-loop effective potential for
the scalar gravitino condensate field $\sigma(x)$ (with vacuum expectation value $\sigma_c \propto \langle
\overline \psi_\mu \,  \psi^\mu \rangle $) has a double-well shape as
a function of $\sigma(x)$ which is symmetric about the origin (\emph{cf.} fig.~\ref{fig:pot2}), as
dictated by the fact that the sign of a fermion mass does not have
physical significance. Dynamical generation of the gravitino mass
occurs at the non-trivial minima corresponding to $\sigma_c  \ne 0$.
The potential of the $\sigma_c$ field is also flat near the origin,
and this has been identified in \cite{emdyno} with a first
inflationary phase.

In \cite{ahm} the one-loop effective potential was derived by first
formulating the theory on a curved de Sitter
background~\cite{fradkin,Fradkin:1983mq}, with cosmological constant (one-loop
induced) $\Lambda > 0$, \emph{not} to be confused with the (negative) tree-level one $\Lambda_0$ (\ref{l0tree}),
and then integrating out spin-2 (graviton) and
spin 3/2 (gravitino) quantum fluctuations in a given class of gauges
(\emph{physical}), before considering the flat limit $\Lambda \to 0$
in a self-consistent way. The detailed analysis in \cite{ahm},
performed in the \emph{physical gauge}, has demonstrated that the
dynamically broken phase is then stable (in the sense of the
effective action not being characterized by imaginary parts)
provided the scale of the gravitino condensate is equal or below the
scale of spontaneous breaking of global SUSY:
\begin{equation}\label{absimparts}
\sigma^2 \le f^2 ~.
\end{equation}
which guarantees the aforementioned result on the necessity of the negative nature of
the tree-level cosmological constant (\ref{l0tree}).

The former result demonstrates the importance of the existence of
global SUSY breaking scale for the stability of the phase where
dynamical generation of gravitino masses occurs, which was not
considered in the previous literature~\cite{odintsov}. In
super-conformal versions of SUGRA, \emph{e.g}. those in
ref.~\cite{confsugra,nmssm}, phenomenologically realistic scales for
$f^2$ and gravitino mass of order of the GUT scale, appear for
appropriate values of the expectation value of the conformal factor.
These imply inflationary scenarios in perfect agreement with the
Planck data~\cite{emdyno,Planck}, on equal footing to the original
Starobinsky model.

In \cite{ahmstaro} we considered an extension of the analysis of
\cite{ahm} to the case where the de Sitter parameter $\Lambda$ is
perturbatively small compared to $m_P^2$, but \emph{non zero}, so
that truncation of the series to order $\Lambda^2$ suffices. This is
in the spirit of the original Starobinsky model~\cite{staro}, with
the r\^ole of matter fulfilled by the now-massive gravitino field.
Specifically, we were interested in the behavior of the effective
potential near the non-trivial minimum, where $\sigma \simeq \sigma_c $ is a
non-zero constant (\emph{cf} Fig. \ref{fig:pot2}). The one-loop
effective potential, obtained by integrating out~\cite{fradkin,Fradkin:1983mq} gravitons and
(massive) gravitino fields in the scalar channel (after appropriate
euclideanisation), may be expressed as a power series in $\Lambda$:
\begin{align}\label{effactionl2}
    \Gamma\simeq S_{\rm cl}-\frac{24\pi^2}{\Lambda^2 }\big(&\alpha^F_0+\alpha_0^B
    + \left(\alpha^F_{1}+ \alpha^B_{1}\right)\Lambda\nonumber\\
    &\qquad+\left(\alpha^F_{2}+ \alpha^B_{2}\right)\Lambda^2+\dots\big)~,
\end{align} where $S_{\rm cl}$ denotes the classical action with tree-level cosmological constant $\Lambda_0$ (to be contrasted with the one-loop cosmological constant $\Lambda$~\footnote{The reader should notice that, upon the restriction (\ref{absimparts}) guaranteeing the absence
of imaginary parts in the one-loop effective action, the tree-level
cosmological constant (\ref{l0tree}) $\Lambda_0 < 0$, while the one-loop one $\Lambda > 0$, as appropriate for a de Sitter background. Thus, $\Lambda_0$ should \emph{not} be confused with the current-epoch positive cosmological constant $\tilde \Lambda_0$, which we introduce later on, in section \ref{sec:rvmpresent}, when we discuss RVM (\emph{cf.} (\ref{l0cdm})).}):
    \begin{align}\label{l0}
        -\frac{1}{2\kappa^2}\int d^4 x \sqrt{g}\left(\widehat{R}-2\Lambda_0\right), \quad
        \Lambda_0=\kappa^2\left(\sigma_c^2-f^2\right)~,
    \end{align}
with $\widehat R$ denoting the fixed $S^4$ background we expand
around ($\widehat R=4\Lambda$, and the 4-dimensional Euclidean
Volume  is $24\pi^2/\Lambda^2$), and the $\alpha$'s indicate the
bosonic (graviton) and fermionic (gravitino) quantum corrections at
each order in $\Lambda$.

The leading order term in $\Lambda$ is then the effective action
found in \cite{ahm} in the limit $\Lambda\to0$,
    \begin{align}
         \Gamma_{\Lambda\to0}\simeq-\frac{24\pi^2}{\Lambda^2}\left(-\frac{\Lambda_0}{\kappa^2}+\alpha_0^F+\alpha_0^B\right)
        \equiv\frac{24\pi^2}{\Lambda^2}\frac{\Lambda_1}{\kappa^2},
    \end{align}
with
    \begin{align} \label{LL1}
\Lambda_{1}=-\, \kappa^2 \,
\left(-\frac{\Lambda_0}{\kappa^2}+\alpha_0^F+\alpha_0^B\right)~,
    \end{align}
 where \begin{align}\label{fermion}
		&\alpha_0^{F} = {\tilde \kappa}^4 \, \sigma_c^4 \, \Big(0.100\,  \ln \left( \frac{{\tilde \kappa}^2 \, \sigma_c^2}{3 \mu ^2}\right) + 0.126 \Big)~,
	\end{align}
and \begin{align}\label{boson}
		\alpha_0^{B}= \kappa ^4 \, \left(f^2-\sigma_c^2\right)^2 \left(0.027 - 0.018 \ln \left(\frac{3 \kappa ^2 \left(f^2-\sigma_c^2\right)}{2 \mu ^2}\right)\right)~,
		\end{align}
indicate the leading (as $\Lambda \to 0$) contributions to the effective potential from bosonic (graviton) and fermionic (gravitino) quantum fluctuations respectively, to one-loop order. Above, $\mu$ is a RG
scale, associated with a short-distance proper time
cutoff~\cite{ahm}, \emph{not} to be confused with the RG scale of the RVM $\mu_c(t)$ (cf. Sect. II),
which is such that the flow from Ultraviolet (UV)
to Infrared  (IR) corresponds to the direction of increasing $\mu$,
$\sigma_c$ denotes the gravitino scalar condensate $\sigma_c \propto \tilde \kappa \,
\langle \overline \psi_\mu \, \psi^\mu \rangle $ at the non-trivial
minimum of the one-loop effective potential (\emph{cf.}
Fig.~\ref{fig:pot2}),  ${\tilde \kappa}$ is the conformally-rescaled gravitational
constant in the Jordan-frame SUGRA model of \cite{confsugra},
defined in (\ref{tildcoupl}), corresponding to
a non-trivial v.e.v. of the conformal (`dilaton') factor, $\langle  \phi \rangle \ne 0$,
assumed to be stabilized by means of an
appropriate potential, leading to the breaking of the conformal
symmetry. In the case of standard ${\mathcal N}=1$ SUGRA, $\langle
\phi \rangle = 0$.

The remaining (higher order in $\Lambda$) one-loop quantum corrections then, proportional to
$\Lambda$ and $\Lambda^2$ may be identified respectively with
Einstein-Hilbert $R$-type and Starobinsky $R^2$-type terms in an
effective action of the form~\footnote{The reader should recall at this stage that the sign of $g$ in $\sqrt{g}$ and the overall minus sign in front of the right-hand-side of (\ref{effactionl3})
is due to the Euclidean-signature
formulation of the path integral and disappears upon analytic continuation back to the Minkowski space-time at the end of the computations, which is necessary in order to make contact with phenomenology/cosmology (see, \emph{e.g}. (\ref{staroaction})). This should be understood in what follows, and especially in the context of linking the SUGRA model with the RVM in section \ref{sec:4}.}
\begin{align}\label{effactionl3}
\Gamma\simeq&-\frac{1}{2\kappa^2} \int d^4 x \sqrt{g}
\left[\left(\widehat R-2\Lambda_1\right)  +\alpha_1 \, \widehat R+
\alpha_2 \, \widehat R^2\right]~,
\end{align}
where we have combined terms of order $\Lambda^2$ into curvature
scalar square terms. For general backgrounds such terms would
correspond to invariants of the form ${\widehat
R}_{\mu\nu\rho\sigma} \, {\widehat R}^{\mu\nu\rho\sigma} $,
${\widehat R}_{\mu\nu} \, {\widehat R}^{\mu\nu}$ and ${\widehat
R}^2$, which for a de Sitter background all combine to yield
${\widehat R}^2$ terms~\footnote{In the pure SUGRA case, with no dilaton frame functions, the fact that the
Gauss Bonnet combination $R_{\mu\nu\rho\sigma} \, {R}^{\mu\nu\rho\sigma} - 4
{R}_{\mu\nu} \, {R}^{\mu\nu} + {R}^2$ is a total derivative in four space-time dimension, implies that one can consider only the Ricci-scalar and Ricci-tensor squared terms as independent. This is not the case though in the conformal SUGRA case~\cite{confsugra}.}. The coefficients $\alpha_1$  and $\alpha_2$
in (\ref{effactionl3}) absorb the non-polynomial (logarithmic) in $\Lambda$ contributions,
so that we may then identify \eqref{effactionl3} with
\eqref{effactionl2} via
    \begin{align}\label{alpha}
         \alpha_1=\frac{\kappa^2}{2}\left(\alpha^F_1+\alpha^B_1\right)~,\quad
        \alpha_2=\frac{\kappa^2}{8}\left(\alpha^F_2+\alpha^B_2\right)~,
    \end{align}
where we note that $\alpha_1$ is dimensionless whereas $\alpha_2$
has dimension of inverse mass squared.
The coefficients $\alpha_i^{F,B}$, $i=1,2$ in Eq.~(\ref{alpha})
can be computed using
the results of \cite{ahm}, derived via an
asymptotic expansion: 
\begin{eqnarray}\label{aif}
\alpha^F_1&=& 0.067\, \tilde\kappa^2 \sigma_c ^2  -0.021\,
\tilde\kappa^2 \sigma_c ^2 \, {\rm ln}
\left(\frac{\Lambda}{\mu^2}\right)
 + \nonumber \\
&& 0.073\, \tilde\kappa^2 \sigma_c ^2 \, {\rm ln}
\left(\frac{\tilde\kappa^2\sigma_c^2}{\mu^2} \right)~,
\nonumber \\
\alpha^F_{2}&=& 0.029 + 0.014\, {\rm ln} \left(\frac{\tilde\kappa^2\sigma_c^2}{\mu^2}\right) - \nonumber \\
&& -0.029\, {\rm ln} \left(\frac{\Lambda}{\mu^2}\right)~,
\end{eqnarray}
and
    \begin{eqnarray}\label{aib}
\alpha^B_1&=& -0.083 \Lambda_0 + 0.018\, \Lambda_0 \, {\rm ln} \left(\frac{\Lambda }{3 \mu ^2}\right)  + \nonumber \\
&&0.049\, \Lambda_0\,  {\rm ln} \left(-\frac{3 \Lambda_0}{\mu
^2}\right)~,
\nonumber \\
\alpha^B_{2}&=& 0.020 +  0.021\, {\rm ln} \left(\frac{\Lambda }{3 \mu ^2}\right) - \nonumber \\
&&0.014\, {\rm ln} \left(-\frac{6 \Lambda_0}{\mu ^2}\right)~.
\end{eqnarray}
To identify the conditions
for phenomenologically acceptable Starobinsky inflation around the
non-trivial minima of the broken SUGRA phase of our model, we
\emph{impose} first the \emph{cancellation} of the ``classical''
Einstein-Hilbert space term $\widehat R $ by the ``cosmological
constant'' term $\Lambda_1$, i.e. that
\begin{equation}\label{condition}
\widehat R = 4 \, \Lambda = 2\, \Lambda_1 \equiv - 2\, \kappa^2 \,
\left(-\frac{\Lambda_0}{\kappa^2}+\alpha_0^F+\alpha_0^B\right) > 0~.
\end{equation}
This condition should be understood as a necessary one
characterizing our background in order to produce
phenomenologically-acceptable Starobinsky inflation in the broken
SUGRA phase following the first inflationary stage, as discussed in
\cite{emdyno}. This may naturally be understood as a generalization
of the relation $\widehat R=2\Lambda_1=0$, imposed in \cite{ahm} as
a self-consistency condition for the dynamical generation of a
gravitino mass in the flat (zero $\Lambda$) limit.

From Eq.~(\ref{condition}) it follows that the (positive) cosmological constant
$\Lambda > 0$ satisfies the four-dimensional Einstein equations in the
non-trivial minimum, and in fact coincides with the value of the
one-loop effective potential of the gravitino condensate at this
minimum. As we discussed in \cite{ahm}, this non-vanishing positive
value of the effective potential is consistent with the generic
features of dynamical breaking of supersymmetry~\cite{witten}. In
terms of the Starobinsky inflationary potential (\ref{staropotent}),
the value $\Lambda > 0$ corresponds to the approximately constant
value of this potential in the high $\varphi$-field regime
($\kappa\vp\gg1$) of Fig.~\ref{fig:potstar}, in the flat region
where Starobinsky-type inflation takes place. Thus we may set
\begin{equation}\label{hubble}
\Lambda \sim 3\, H_I^2
\end{equation}
where $H_I$ the (approximately) constant Hubble scale during
inflation, which is constrained by the current data to satisfy
(\ref{upperH}) or (\ref{upperHbic}). In the SUGRA context under discussion $H_I$ is linked to the scale of global SUSY breaking through $H_I\simeq f/M_{Pl}$.

The effective Newton's constant in  (\ref{effactionl3}), after the imposition of (\ref{condition}), is then defined as
\begin{equation}\label{keff}
\kappa_{\rm eff}^2=\frac{\kappa^2}{\alpha_1}~,
\end{equation}
and from this, we can
express the effective Starobinsky parameter (\ref{staroaction}) in
terms of $\kappa_{\rm eff}$ as
\begin{equation}\label{beff}
\beta_{\rm eff} \equiv
\frac{\alpha_2}{\alpha_1}~.
\end{equation}
This condition thus makes a direct link between
the action (\ref{effactionl2}) with a Starobinsky type action
(\ref{staroaction}). Comparing with (\ref{staroaction}) we can
determine the effective scalaron mass in this case:
\begin{equation}\label{staroours}
{\mathcal M} = \sqrt{\frac{8 \pi}{3} \, \frac{\alpha_1}{\alpha_2}
}~,
\end{equation}
As we know, this mass parameter also sets the order of magnitude of
the inflationary scale in the Starobinsky model.

We may then determine the coefficients $\alpha_1$ and $\alpha_2$ in
order to evaluate the scale $1/\sqrt{\beta}\sim{\cal M}$ of the
effective Starobinsky potential given in Fig.~\ref{fig:potstar} in
this case, and thus the scale of the second inflationary phase.

In \cite{ahmstaro} we searched numerically for points in the parameter space such that:
\begin{itemize}
\item{} The effective equations
    \begin{align}\label{effeq}
         \frac{\partial\Gamma}{\partial\Lambda}=0~, \quad
         \frac{\partial\Gamma}{\partial\sigma}\Big|_{\sigma=\sigma_c}=0~,
    \end{align}
are satisfied, together with the condition (\ref{condition}),

\item{} the cosmological constant $\Lambda$ is small and positive,
satisfying (\ref{hubble}), and for phenomenological reasons
it should be of order
\begin{equation}\label{L1}
0<\Lambda \sim 10^{-10}M^2_{\rm Pl}~,
\end{equation}
to ensure the validity of our expansion in $\Lambda$, consistent
with the phenomenology of Planck- satellite data~\cite{Planck}.

\item{} The scalaron mass should also be of order ${\mathcal M} \sim 10^{-5} \, M_{\rm Pl} $,
hence allowing us to achieve phenomenologically acceptable
Starobinsky inflation in the massive gravitino phase, consistent
with the Planck-satellite data~\cite{Planck}.

\end{itemize}

For $\tilde \kappa=\kappa$ (\emph{i.e}. for non-conformal
supergravity), we were unable to find any solutions satisfying these
constraints. This of course may not be surprising, given the
previously demonstrated non-phenomenological suitability of this
simple model \cite{ahm}. If we consider $\tilde \kappa \gg \kappa$
however, we find that we are able to satisfy the above constraints
for a range of values~\footnote{A comment concerning SUGRA models in
the Jordan frame with such large values for their frame functions is
in order here. In our approach, the dilaton $2\phi$ could be a
genuine (dimensionless) dilation scalar field arising in the gravitational multiplet of string theory, whose
low-energy limit may be identified with some form of SUGRA action.
In our normalization the string coupling would be $g_s \equiv e^\phi
= \Big({\tilde \kappa}/\kappa\Big)^{-1/2}$. In such a case, a value of $\tilde \kappa
= e^{-2\langle \phi \rangle}\, \kappa = {\mathcal O}(10^{3-4})\, \kappa$
would imply a large negative v.e.v. of the (four-dimensional)
dilaton field of order $\langle \phi \rangle = -{\mathcal O}(5) <
0$, and thus a weak string coupling squared $g_s = {\mathcal
O}(10^{-2})$, which may not be far from values attained in realistic
phenomenological string models. On the other hand, in the
Jordan-frame SUGRA models of \cite{confsugra}, the frame function
reads $\Phi \equiv e^{-2\phi} =  1 - \frac{1}{3}\Big(S{\overline
S} + \sum_{u,d} \, H_i H^\dagger_i \Big) - \frac{1}{2} \chi \,
\Big(-H_u^0 \, H_d^0 + H_u^+ \, H_d^- + {\rm h.c.} \Big)$, in the
notation of \cite{nmssm} for the various matter super fields of the
next-to-minimal supersymmetric standard model that can be embedded
in such supergravities. The quantity $\chi$ is a constant parameter.
At energy scales much lower than GUT, it is expected that the
various fields take on subplanckian values, in which case the frame
function is almost one, and hence $\tilde \kappa \simeq \kappa $ for
such models today. To ensure $\tilde \kappa \gg \kappa$,  and thus
large values of the frame function, $\Phi \gg 1$, as required in our
analysis, one needs to invoke trasnplanckian values for some of the
fields, $H_{u,d}^0$, and large values of $\chi$, which may indeed
characterize the inflationary phase of such theories. A similar
situation occurs for the values of the Higgs field (playing the role
of the inflaton) in the non-supersymmetric Higgs inflation
models~\cite{higgsinfl}.}.

In general, typical values obtained in phenomenological realistic conformal SUGRA models
 satisfy
$\tilde \kappa \gg \kappa$ (\emph{e.g.} of order $\tilde \kappa / \kappa = {\mathcal O}(10^{3} - 10^4))$, under the constraints (\ref{hubble}) and (\ref{L1}), in such a way that:
 {\small  \begin{equation}\label{scales}
  \Lambda \sim 3H_I^2 \sim m_{3/2}^2 \sim {\tilde \kappa}^2 \sigma_c^2 \sim \kappa^2 f^2 \ll \mu^2 = 8\pi/\kappa^2\,, \sigma_c^2 \ll f^2.
  \end{equation}} Since the scale of SUSY breaking must be in the ballpark of the typical GUT scale associated to the inflation, namely $\sqrt{f}\sim 10^{16}$ GeV$\sim 10^{-2}\,M_{Pl}$, from the above we have  $\Lambda \sim \kappa^2f^2=f^2/M_{Pl}^2\sim 10^{27}$ GeV$^2$. As a result the scale of the gravitino is some two to three orders of magnitude below the GUT scale, that is to say,  $m_{3/2}\sim \sqrt{\Lambda}\gtrsim 10^{13}$ GeV$\sim 10^{-5}\,M_{Pl}$.
These values are compatible with both the combined Planck and Bicep2 bound (\ref{upperHbic})
and the typical mass of the gravitino in this framework ~\cite{ahmstaro}.

Exit from the inflationary phase is, of course, a complicated issue
which we shall not discuss here at the level of the SUGRA model
itself, aside from the observation that it can be achieved by
coherent oscillations of the gravitino condensate field around its
minima and subsequent decays to radiation and matter fields (thus requiring detailed knowledge of the matter content of the SUGRA models in order to arrive at quantitative predictions for the exit phase), or tunnelling processes \`a  la Vilenkin~\cite{vilenkin}.
However, in the next section we will show that the SUGRA model can
be represented by an effective running vacuum model along the lines
indicated in Sect.\ref{sec:running}, and from this point of view the
exiting from the inflationary phase into the standard radiation
phase can be guaranteed on very general grounds.

Before doing so, though, we should make some important remarks concerning the presence of logarithms of the de Sitter scale $\Lambda$ in the coefficients $\alpha_i$ of the curvature terms of the effective action (\ref{effactionl3}).
When one computes the effective action in a fixed de Sitter background, it is tempting to identify a $\Lambda$ term with the Ricci scalar, which eventually will be allowed to depend on time. Thus, naively, the presence of logarithms would imply non-polynomial terms of the form $R\, {\rm ln}R $ which would be problematic for any RVM interpretation of the exit from the inflationary phase, as it would contradict the spirit of the approach where only integer powers of the curvature terms would be allowed in the respective flow equations~\cite{ShapSol,SolStef05,Fossil07,JSPRev2013,SolGo2015a}.
Fortunately this is not the case. To understand this, we first remark that any effective action obtained by integrating out massive degrees of freedom such as gravitino fields, that we restrict ourselves here, must consists for reasons of covariance and consistency of the weak gravitational fluctuations about the de Sitter background only of polynomial structures of the curvature tensors, for instance to fourth order in derivatives terms involving the squares of the Ricci scalar and Ricci tensors and covariant derivatives thereof. Any $R\, {\rm ln}R$ term would be incompatible with the weak gravity perturbative expansion
about a background, say of constant non-zero curvature.

Thus, the coefficients $\alpha_1$ and $\alpha_2$ in the action (\ref{effactionl3}) are kept \emph{fixed}, not undergoing temporal evolution, which is guaranteed by the fixing of the two free scales in the problem $\mu$ (\ref{scales}) and $\Lambda$ (\ref{hubble}).
Notice that the scale $\mu$ should not be confused with the subsequent RG scale $\mu_c (t)$ that describes the cosmological evolution of the RVM vacuum (cf. Sect.\,\ref{sec:running}).
Indeed, the scale $\mu$ first of all is a high energy cut-off. As already mentioned, it plays the role of a proper-time cutoff scale~\cite{ahm}, appearing in the integral representations of some $\zeta$-functions that are part of the determinants arising in the path integral of the SUGRA action arising from integrating out massive spin 3/2 (gravitino) and spin 2 (graviton) fluctuations about the de Sitter background. The scale $\mu$ is therefore, in contrast to $\mu_c(t)$, an inverse renormalization group scale. Its value has to be fixed so as to guarantee SUGRA breaking and to generate a fixed gravitino mass which should not depend on time. This implies that the spontaneous breaking of SUGRA and the inflationary phase are characterised by such fixed scales, which implies the time independence of $\Lambda$ (or, equivalently, the Hubble parameter) during inflation, the gravitino mass, related to the gravitino condensate vacuum expectation value $\sigma_c$, and thus the coefficients $\alpha_1$ and $\alpha_2$. On the other hand,
integer positive powers of $\Lambda$, appearing in the effective action may be replaced by higher order tensorial structures involving the square of the curvature tensors, which are allowed to vary with the cosmic time during the RVM phase after exit from inflation.
Notice that microscopically the exit phase is characterised by an unknown sort of phase transition, either through decays of the gravitino condensates to
matter parts and reheating of the inflated universe, or tunnelling, as mentioned previously, and thus using different RG running to relate various eras of the Universe after inflation is to be expected.

\vspace{0.2cm}
\section{``Decay'' of Effective Vacuum Energy: running vacuum model \label{sec:4}}

The main aim of this section  is to demonstrate that there exists a
family of time-dependent effective vacuum energy decaying models of
\emph{running type}, i.e. the class of the running vacuum models
(RVM's) introduced in  Sect.\,\ref{sec:running}, which characterize
the evolution of the Universe from the exit of the Starobinsky
inflationary phase till the present era. In fact, the RVM's are able
to interpolate on very general grounds the primeval de Sitter epoch
with the late time de Sitter era, i.e. the dark energy one, where a
much smaller cosmological constant essentially dominates. We shall
follow the approach of the RVM outlined in Sect.\,\ref{sec:running},
in which the vacuum energy density $\rho_{\Lambda}(H)$ varies with
time through its dependence on $H=H(t)$. The Hubble parameter,
having dimension of energy in natural units, acts as the natural
running scale via the RG equation Eq.(\ref{runningrho}). As
mentioned in Sect.\,\ref{sec:running}, only the even powers of $H$
can be involved in that equation, owing to the general covariance of
the effective action. This is an important point to make possible a
general QFT description of this RG approach and is essential for the
connection with the SUGRA model under discussion.

It is evident that the expansion (\ref{runningrho}) quickly
converges at low energies, where $H$ is rather small -- certainly
much smaller than any particle mass. No other $H^{2n}$-term beyond
$H^2$ (not even $H^4$) can contribute significantly on the
\textit{r.h.s.} of equation (\ref{runningrho}) at any stage of the
cosmological history below the GUT scale $M_X\sim 10^{16}$ GeV,
where presumably inflation occurs.

On the other hand, if we want to deal with the physics of inflation
and in general to the very early states of the cosmic evolution, we
have to keep at least the term $H^4$, which in fact is the dominant
term in the series (\ref{runningrho}) during the high energy regime.
In contrast, the terms $H^6/M_i^2$ and above are less and less
important because these higher and higher powers of $H$ are
suppressed by the inverse powers of the heavy fermion and boson
masses in the GUT, as required by the Appelquist-Carazzone
decoupling theorem. Therefore, the dominant part of the series
(\ref{runningrho}) is expected to be naturally truncated at the
$H^4$ term. Higher order terms should contain the bulk of the high
energy contributions within Quantum Field Theory in curved
spacetime, namely within a semi-classical description of gravity
near but (possibly a few orders) below the Planck scale. Models of
inflation based on higher order terms inspired by the RG framework
exist since long in the literature (see \cite{ShapSol}) as well as
the unified inflation-dark energy framework of \cite{Fossil07}. For
a more phenomenological treatment unrelated to the RG, see
\cite{LM94,LT96,ML2000,CT06}.

\subsection{A distinct class of running vacuum models}\label{ref:class}

Based on the above arguments, it is natural to consider the case in
which the highest power of the
Hubble rate in the RG Eq.\,(\ref{runningrho}) is $H^4$.
Integrating the RG equation provides the simplest realization of RVM that
can describe inflation and the various stages of the FLRW regime:
\begin{equation}\label{lambda}
\rL(H) = \frac{\Lambda(H)}{\kappa^2}=\frac{3}{\kappa^2}\left(c_0 + \nu H^{2} + \alpha
\frac{H^{4}}{H_{I}^{2}}\right) \;.
\end{equation}
Here $c_0$ is an integration constant (with dimension $+2$ in
natural units, i.e. energy squared) which can be fixed
from the low energy data of the
current universe \cite{BPS09,GoSolBas2015}.
On the other hand the
dimensionless coefficients are given as follows:
\begin{equation}\label{eq:nuloopcoeff}
\nu=\frac{1}{48\pi^2}\, \sum_{i=F,B} a_i\frac{M_i^2}{M_{\rm
Pl}^2}\,,
\end{equation}
and
\begin{equation}\label{eq:alphaloopcoeff}
\alpha=\frac{1}{96\pi^2}\, \frac{H_I^2}{M_{\rm Pl}^2}\sum_{i=F,B}
b_i\,.
\end{equation}

At this point, we would like to make some comments which will
hopefully make the reader appreciate the physical interpretation of
the running vacuum scenario. The coefficient $\nu$ behaves as the
reduced (dimensionless) beta-function for the RG running of
$\rho_{\Lambda}$ at low energies, whereas $\alpha$ plays a similar
role at high energies. Notice that the index $i$ depends on whether
bosons ($B$) or fermions ($F$) dominate in the loop contributions.
Of course, since the coefficients $(\nu,\alpha)$ play the role of
one-loop beta-functions (at the respective low and high energy
scales) they are expected to be naturally small because $M_i^2\ll
M_{\rm Pl}^2$ for all the particles, even for the heavy fields of a
typical GUT. Indeed, an estimate of $\nu$ within a generic GUT is
found in the range $|\nu|=10^{-6}-10^{-3}$ \cite{Fossil07}. The
dimensionless coefficient $\alpha$ is also small, $|\alpha|\ll 1$,
because the inflationary scale $H_I$ is certainly below the Planck
scale, see Eq.\,(\ref{upperHbic}).
From the observational viewpoint, utilizing a joint likelihood
analysis of the recent supernovae type Ia data, the CMB shift
parameter, and the Baryonic Acoustic Oscillations, it has been found
$|\nu|={\cal
O}(10^{-3})$\,\cite{BPS09,GrandeET11,GoSolBas2015,GoSol2015}, which
is nicely in accordance with the aforementioned theoretical expectations
as well as it insures a mild dynamical behavior of the vacuum energy
at low energies.
As we have already stated in section  \ref{sec:running}, the Quantum-gravity
corrections in the Starobinsky model have been found in the context
of the RG analysis \cite{copeland,litim} through the appropriate
beta-functions. The fact that the nature of the main coefficients of
both theories (running vacuum and Starobinsky) is based on the RG
approach is a hint that perhaps there is a possible connection
between the two models.

Indeed as we will confirm below, this is the case. In particular,
let us start with the effective action (\ref{effactionl3}), with
coefficients (\ref{alpha}),  (\ref{aif}), (\ref{aib}) and the
constraints (\ref{condition}), (\ref{scales}). As we have seen,
this action was obtained after integrating out both quantum-gravity
(metric) fluctuations and massive gravitino fields. The action
admits dynamical solutions of de-Sitter vacua with small
cosmological constant $\Lambda$ of order less than the GUT scale at
early epochs of the Universe. At the non-trivial minimum, the
effective potential takes on a value of order $\Lambda$ (\emph{cf}.
Fig.~\ref{fig:pot2}). Around the minimum, we
should replace $\Lambda$ by an effective Hubble parameter during
inflation, $H_I$, Eq.~(\ref{hubble}).

At the end of the day the effective (dynamical) vacuum energy density, $\rho_\Lambda(H)$,
during the inflationary phase of our SUGRA model can be extracted from the SUGRA effective action $\Gamma$ \,(\ref{effactionl3}),
upon applying the constraint (\ref{condition}) and analytically continuing the results back to Minkowski space-time signature.
In particular, the effective potential is defined as $V_{\rm eff} \equiv - \Gamma \rightarrow \int d^4 x \sqrt{-g}\, \rho_\Lambda (H)$.
Doing so, we observe that the so obtained $\rho_\Lambda (H)$, remarkably, adopts precisely the generic RVM structure (\ref{lambda}) around that phase, in which the Ricci scalar --  see Eq.\,(\ref{SF.3b}) below ---  boils down to
$R\simeq 12H^{2}$ since $H$ remains (approximately) constant in this phase.

Some  important remarks are in order here. The imposition of the constraint (\ref{condition}) during the Starobinsky inflationary phase implies, as already mentioned, that the correct phenomenology is attained as a result of the effective gravitational coupling
(\ref{keff}) that characterises that phase.
So, if the constraint \emph{was} an \emph{exact result}, the
effective vacuum energy density of the SUGRA model
would then correspond to the ${\widehat R}^2 \rightarrow 144 \, H^4$ terms in (\ref{effactionl3})
with (\ref{keff}) playing the r\^ole of the effective gravitational constant,
{ \begin{equation}\label{rholambda1}
\rho_\Lambda^{\rm SUGRA} (H)^{\rm exact}_{\rm constraint} = \frac{72}{\kappa_{\rm eff}^2} \, \frac{\alpha_2}{\alpha_1} \,  H^4 =
\frac{18}{\kappa_{\rm eff}^2} \, \frac{\alpha_2^F + \alpha_2^B}{\alpha_1^F + \alpha_1^B} \,  H^4~,\end{equation}}
where we used (\ref{alpha}), (\ref{aif}), (\ref{aib}).
The form (\ref{rholambda1}) constitutes an admissible class of RVM (\emph{cf.} Eq.~(\ref{lambda})).
Notice that in Eq.~(\ref{rholambda1}) there is no $\nu$ term. This is important, in the sense that in such a model, as a result of the effective gravitational constant (\ref{keff}) entering the game, which in this scenario~\cite{ahmstaro} is viewed as the `\emph{physica}l'  reduced Planck mass of order $10^{18}$
~GeV, the gravitino  mass and global SUSY breaking scales,
(\ref{scales}), when expressed in terms of $\kappa_{\rm eff}$ are of order one, that is one encounters a Planck-scale gravitino. Despite this, the vanishing of $\nu$ makes the renormalization-group equation (\ref{rholambda1}) a consistent one within the perturbative class of (\ref{lambda}).

However the above construction leads to the absence
of a present-era (small, positive) cosmological constant $c_0$. This arises from the fact that we imposed the constraint
(\ref{condition}) exactly. It may well be that such a condition leaves (non-perturbatively, when all the higher than one-loop contributions are taken into account) a very small (constant in cosmic time) contribution $c_0 > 0$ which is preserved until the present day.  Unfortunately, our one-loop construction does not allow us to explain the magnitude and the sign of this constant term, but this is equivalent to offering a solution to the cosmological constant problem, which of course our approximate one-loop analysis cannot provide. While we do not have a quantitative calculation at this point, the above argument provides at least an interesting qualitative explanation, to wit: the origin of the current cosmological term $\rho_\Lambda^0$ in the model might well be attributed to quantum (non-perturbative) effects in the SUGRA effective action, which prevent the complete cancellation (\ref{condition}) from being realised. The constant residue $c_0$ is then transferred throughout the cosmic history and pops up in our days in the form of the tiny vacuum energy $\rho_{\Lambda}^0=(3/\kappa^2)(c_0+\nu H_0^2)\simeq 3c_0/\kappa^2$.

Under this assumption, then, the initial gravitational coupling $\kappa$, and thus the Einstein term $\frac{1}{2\kappa^2} \int d^4 x \sqrt{-g}\, \widehat R$ would enter the game during the exit phase from inflation~\footnote{The reader should bear in mind that, since during the inflationary phase the scalar degree of freedom of the Starobinsky action is slowly rolling, if there is inflation in the conformally rescaled metric (\ref{confmetric}), there is also inflation in the initial metric. The Starobinsky inflation arguments are also not affected if a small contribution to the cosmological constant, of order of the present-era one, enters the effective action (\ref{steps}), as this is negligible compared to the Hubble scale of inflation.}.
In such a case, in the exit phase, the effective vacuum energy of the SUGRA  model at the inflationary phase should correspond to \emph{both} $\alpha_1$ and $\alpha_2$ terms of (\ref{effactionl3}), with the constraint (\ref{condition}) failing by a tiny amount $\tilde c_0 > 0$ corresponding to the present-era cosmological constant.
\begin{equation}\label{rholambda}
\rho_\Lambda^{\rm SUGRA} (H) = \frac{1}{\kappa^2} \Big( \tilde c_0 + 6 \alpha_1 \, H^2 + 72\, \alpha_2\,  H^4 \Big)
\end{equation}
where the explicit form of the coefficients $\alpha_i$, $i=1,2$, given by Eqs.~(\ref{alpha}), (\ref{aif}) and (\ref{aib}), in which the scales $\mu$ and $\Lambda$ are \emph{fixed} through (\ref{hubble}) and (\ref{scales}) respectively. As mentioned already, fixing of the scale $\mu$ and $\Lambda$, implies fixed values for the gravitino mass.

It is worth stressing that the aforementioned ambiguity concerning the failure of the exact constraint (\ref{condition}) can be avoided altogether by observing  that the corresponding $\rho_{\Lambda}(H)$ ultimately derives from the RG equation (\ref{runningrho}) discussed in Sect.\,\ref{sec:running}. It is therefore more appropriate, and elegant, if one performs the matching between  the running vacuum energies  $\rho_\Lambda (H)$ in the SUGRA model and RVM by equating the corresponding ``RG beta'' functions,  $\frac{d\, \rho_\Lambda}{d \, {\rm ln}\, H^2}$:
\begin{eqnarray}\label{rhoder}
&&\frac{d\, \rho_\Lambda}{d \, {\rm ln}\, H^2}  = 6\, \kappa^{-2} \, \alpha_1 \, H^2 + 144\, \kappa^{-2}\,  \alpha_2 \, H^4 \simeq
  \nonumber \\  && 3\, ( \alpha_1^F + \alpha_1^B ) \, H^2 + 18 \, (\alpha_2^F  + \alpha_2^B) \, H^4
  \end{eqnarray}
 It is remarkable that the effective dynamical vacuum energy density $\rL(H)$ associated to the SUGRA model under consideration turns out to  follow the general RG of the RVM, see Eq.\,(\ref{runningrho}), in
which the coefficients of the $H^2$ and $H^4$ terms can be computed precisely from the underlying SUGRA framework.

From Eq.\,(\ref{rhoder}) it follows that $\rL=\rL(H)$ evolves
(``runs'') with the (time-evolving) value of $H$. Such evolution
will be studied in more detail in the next section, but is
relatively small in the beginning, namely the varying $H$ is only
slightly below the initial value given in Eq.\,(\ref{hubble}). As a
result the Universe can start with an initial inflationary phase,
which is dominated by the $\sim H^4$ term of (\ref{rhoder}).
However, well after the inflationary period the $\sim H^2$-term
takes over and remains in force until the present time, thereby
providing a mild evolution of the current cosmological ``constant''.

Eventually one has to add to the effective action loop contributions
from other matter fields, including particle multiplicities, but
at the moment we take a mass of order of the gravitino mass and shall comment on the possible additional effects below. The value of $m_{3/2}$ is of order of the GUT
scale, since it is
proportional to the gravitino condensate through $\sigma_c \sim
{\tilde \kappa}^{-1} \, m_{3/2}$, the latter being bounded from
above by the GUT scale: $\sqrt{\sigma_c}\leq \sqrt{f} \sim
10^{-2}M_{\rm Pl}$ [see Eq.\,(\ref{absimparts})].
We must also keep in mind that for phenomenologically acceptable
solutions of the broken SUGRA model the ratio $r\equiv{\tilde
\kappa} /\kappa$  is forced to stay in the range $r={\mathcal
O}(10^3 - 10^4)$.

From the generic values (\ref{scales}) we adopt here we find:
  \begin{eqnarray}\label{betavalue}
\frac{d\, \rho_\Lambda}{d \, {\rm ln}\, H^2}
  \sim 1.59 \, {\tilde \kappa}^2 \sigma_c^2 \, H^2  + 25.76 \, H^4 \;.
\end{eqnarray}
The integrated form of Eq.(\ref{betavalue}), yields of course the effective vacuum energy density at the scale $H$,  $\rL(H)$,  Eq.~(\ref{rholambda}),
 within the current SUGRA scenario.  We thus have:
 \begin{equation}\label{SUGRAL}
\rL(H)\simeq \tilde {c}_{0}+ 1.59\, {\tilde \kappa}^{2}
\sigma^{2}_{c}H^{2}+ 12.88\, H^{4} \;,
\end{equation}
where $\tilde{c}_{0}$ is the integration constant, which will play the r\^ole of the current-era cosmological constant, as already mentioned. The result
naturally adopts the generic form of the canonical RVM,
Eq.\,(\ref{lambda}) with $c_0=\kappa^2\tilde{c}_0/3$ and  the
effective values for the coefficients $\nu$ and $\alpha$ given by
\begin{eqnarray}
\nu_{\rm eff}&\simeq& 0.53 \, \kappa^2\tilde{\kappa}^2\sigma_c^2\simeq {\cal O}\left(\frac{m^2_{3/2}}{M^2_{\rm Pl}}\right)\,,\label{eq:nueff}\\
\alpha_{\rm eff}&\simeq& 4.30\,H_I^2\,\kappa^2\simeq {\cal
O}\left(\frac{H^2_I}{M^2_{\rm Pl}}\right)\label{eq:alphaeff}\,.
\end{eqnarray}
Thus we observe that, within the context of the pure SUGRA model, where only the gravitino plays the r\^ole of ``matter'', both coefficients are small, of typical order $10^{-9}$, in accordance with their interpretation
as $\beta$-function coefficients of the running vacuum energy
density. Let us also note that the above estimate for $\nu_{\rm
eff}$ nicely fits with the formal expression obtained in the
different context of anomaly-induced inflation, where it also takes the structure (\ref{eq:nuloopcoeff}), namely a quantity proportional to
the (squared) ratio of a heavy particle scale (in general a
collection of them) to the Planck mass --  see \,\cite{Fossil07} for
details. In the full SUGRA case, after its coupling to ordinary matter and radiation fields, other SUSY heavy fermions with masses near the SUSY breaking scale should also contribute \phantom{} to $\nu_{\rm eff}$ and $\alpha$, and
this should enhance the value of this coefficient very significantly, thus bringing the obtained
result even closer to the situation studied in Ref.\,\cite{Fossil07}. Overall, such considerations
in phenomenologically realistic SUGRA situations could bring \phantom{} these parameters to a
range $\sim 10^{-4}$ accessible to current observations\,\cite{GoSolBas2015,GoSol2015}.

We next remark that, in the general case where the parameters of the SUGRA model are varied
from the generic values considered in (\ref{scales}), but
within the allowed range, the values of $\nu_{\rm eff}$ and
$\alpha_{\rm eff}$ can also undergo some variation and the sign of
$\nu_{\rm eff}$ could change. However we stress that the sign of
$\alpha_{\rm eff}$ remains always positive, which is essential for a
correct description of inflation. This can be seen explicitly by comparing the
various logarithms involved in the structure of the coefficient $\alpha_2$ of the $H^4$-term in
Eq.\,(\ref{rhoder}), together with the size and sign of their
respective numerical coefficients (\emph{cf.} Eqs.~(\ref{alpha}), (\ref{aif}), (\ref{aib})).
The positivity of $\alpha_2$ is maintained throughout the
physically allowed parameter space, and
derives essentially from the fact that  $H_I^2\sim \tilde{\kappa}^2\, \sigma_c^2\sim \kappa^2\,m_{3/2}^2$,
in agreement with the generic result (\ref{scales}).

Finally, let us note that the circumstance that $\nu_{\rm
eff}$ could have either sign can only affects the dynamics of the
vacuum energy in the late universe. The phenomenological
implications for both signs have actually been explored recently in
\cite{GoSolBas2015,GoSol2015}, see also\,\cite{SolGo2015a}.

The upshot of the above considerations points to the existence of a
remarkable relation between the running vacuum model
Eq.(\ref{lambda}) with that of SUGRA Eq.(\ref{SUGRAL}). In the next
section we discuss the predicted inflationary scenario
 \,\cite{LBS2013,BLS2013} in the context of the general RVM\,\cite{JSPRev2013,SolGo2015a},
and  provide some interesting phenomenology that can be tested for the low energy regime, namely for the current Universe.

\subsection{Running Vacuum Evolution: from current to inflationary era \label{sec:rvmpresent}}
In this section  we investigate the conditions under which the
running vacuum model can provide an inflationary era. The point of this session is first to demonstrate  that
is, if one starts from an inflationary era, at an early epoch, obtained in the  context of a microscopic model, such as the Starobinsky inflation induced in the SUGRA model,
then the RVM can smoothly connect it with the current era, characterised by a
very small value of the vacuum energy, with a cosmology of $\CC$CDM type.
We shall follow a ``\emph{bottom-up}'' approach, in which, by starting from a late epoch FLRW Universe and applying RVM evolution (``backwards'' in cosmic time, or, in a RG sense, an IR to UV flow), one arrives at an inflationary era in the early Universe.
As we shall see, however,  in this bottom-up approach, there is \emph{no unique} way to identify the underlying microscopic model during the de Sitter era, which was to be expected in view of the rather generic features encapsulated in the RVM evolution.

To this end, let us first reproduce the
Friedmann equations in the framework of a running $\rho_{\Lambda}$.
The resulting equations are expected to be formally equivalent to
the $\CC$CDM case, inasmuch as the Cosmological Principle, which is
embedded in the FLRW metric, perfectly allows the possibility of a
time-evolving cosmological term. In general, the Einstein-Hilbert
action is given by (here and in what follows we are back in Minkowski-signature space-time, described by a metric $g_{\mu\nu}$):
\begin{equation}
S_{R,\Lambda}=\int d^{4}x\sqrt{-g}\left[
\frac{1}{2\kappa^{2}}(R-2\Lambda)
+\mathcal{L}_{m}\right]  \label{action1}%
\end{equation}
where in our case $\rho_{\Lambda}(t)=\Lambda (t)/\kappa^2$ represents the effective vacuum energy, which is allowed to vary with the
cosmic time (more specifically as a function of a dynamical
cosmological variable that evolves with time), and $\mathcal{L}_{m}$
is the Lagrangian of matter. Varying the action (\ref{action1}) with
respect to the metric we arrive at
\begin{equation}
R_{\mu \nu }-\frac{1}{2}g_{\mu \nu }R=\kappa^2\,  \tilde{T}_{\mu\nu}\,,
\label{EE}
\end{equation}
where the total $\tilde{T}_{\mu\nu}$ is given by
$\tilde{T}_{\mu\nu}\equiv T_{\mu\nu}-g_{\mu\nu}\,\rho_{\Lambda} $,
with $T_{\mu\nu}=-2\partial{\cal L}_{m}/\partial
g_{\mu\nu}+g_{\mu\nu}\,{\cal L}_{m}$ the energy-momentum tensor
corresponding to the  matter Lagrangian. The extra piece is
$\rho_{\Lambda}=\Lambda/\kappa^2$, that is to say, the vacuum energy
density associated to the presence of $\Lambda(t)$ (with pressure
$p_{\Lambda}=-\rho_{\Lambda}$). Let us remark that this equation of
state (EoS) does not depend on whether the vacuum is dynamical or
not. In contrast to other forms of dark energy, the vacuum is
defined as that for which the EoS parameter is precisely
$\omega=-1$ in any circumstance.

Modeling the expanding universe as a perfect fluid with velocity
$4$-vector field $U_{\mu}$, we obtain
$T_{\mu\nu}=p_{m}\,g_{\mu\nu}+(\rho_{m}+p_{m})\,U_{\mu}U_{\nu}$,
where $\rho_{m}$ is the density of matter-radiation and
$p_{m}=\omega_{m} \rho_{m}$ is the corresponding pressure, in which
$\omega_m$ is the EoS of matter. Obviously, $\tilde{T}_{\mu\nu}$
takes the same form as ${T}_{\mu\nu}$ with $\rho_{\rm
tot}=\rho_{m}+\rho_{\Lambda}$ and $p_{\rm
tot}=p_{m}+p_{\Lambda}=p_{m}-\rho_{\Lambda}$, that is,
$\tilde{T}_{\mu\nu}=
(p_{m}-\rho_{\Lambda})\,g_{\mu\nu}+(\rho_{m}+p_{m})U_{\mu}U_{\nu}$.

In the context of a spatially flat FLRW metric, we derive the
Friedmann equations in the presence of a dynamical $\CC$-term: \be
 \kappa^{2}\rho_{\rm tot}=\kappa^2 \rho_{m}+\Lambda = 3H^2 \;,
\label{friedr} \ee
\be
\kappa^{2}p_{\rm tot}=\kappa^2 p_{m}-\Lambda =-2{\dot H}-3H^2
\label{friedr2} \ee and the Ricci scalar
\begin{equation}
\label{SF.3b} R=g^{\mu\nu}R_{\mu\nu}= 6(2H^{2}+\dot{H}) \;,
\end{equation}
where the overdot denotes derivative with respect to cosmic time
$t$. Note, that the Bianchi identities
$\bigtriangledown^{\mu}\,{\tilde{T}}_{\mu\nu}=0$, insure the
covariance of the theory and, if the Newtonian coupling is strictly
$G=$const.,  entail an energy exchange between vacuum and matter.
\begin{equation}
\dot{\rho}_{m}+3(1+\omega_{m})H\rho_{m}=-\dot{\rho_{\Lambda}}\,.
\label{frie33}
\end{equation}

Combining equations (\ref{friedr}), (\ref{friedr2}) and
(\ref{frie33}), we infer the basic differential equation that
governs the dynamics of the Universe, namely the equation for the
Hubble rate:
\begin{equation}
\dot{H}+\frac{3}{2}(1+\omega_{m})
H^{2}=\frac12\,\kappa^2(1+\omega_{m})\rho_{\Lambda}=
\frac{(1+\omega_{m})\Lambda}{2}\,. \label{frie34}
\end{equation}
Inserting in it the expression (\ref{lambda}) for the dynamical
vacuum energy we arrive at the following equation:
\begin{equation}
\label{HE} \dot
H+\frac{3}{2}(1+\omega_m)H^2\left[1-\nu-\frac{c_0}{H^2}-\alpha\frac{H^2}{H_I^2}\right]=0\,
\end{equation}
The dynamics of this model has been thoroughly discussed in
\cite{LBS2013,BLS2013}, see also \cite{SolGo2015a}. We can summarize it as follows. First of all
we identify  the presence of an inflationary epoch (de Sitter phase)
associated to the constant value solution $H^2=(1-\nu)H_I^2/\alpha$
of Eq.\,(\ref{HE}), which is valid for the very early epoch of
the universe (in which we can neglect $c_0/H^2\ll 1$). In this
regime, solving Eq.(\ref{HE}) we find
\begin{equation}\label{HS1}
 H(a)=\left(\frac{1-\nu}{\alpha}\right)^{1/2}\,\frac{H_{I}}{\sqrt{D\,a^{3(1-\nu)(1+\omega_m)}+1}}\,,
\end{equation}
where $D$ is a positive constant of integration. For the early
universe we assume that matter is essentially relativistic, thus we
take $\omega_m=1/3$ at this point. Overall, one can see from (\ref{HS1}) that for $Da^{4(1-\nu)} \ll 1$ the universe starts
from an unstable inflationary phase [early de Sitter era,
$H^2=(1-\nu) H_I^2/\alpha$] powered by the huge value $H_I$
presumably connected to the scale of a Grand Unified Theory (GUT).
Well after the primeval inflationary era, specifically for $Da^{4(1-\nu)} \gg 1$, the Universe enters the standard radiation phase. Subsequently the radiation component becomes subdominant and
the matter dominated era appears. This is confirmed from the evolution of the vacuum energy and radiation energy densities. If we neglect $\nu$ and $c_0/H^2$ in this early epoch, which is justified, we can insert (\ref{HS1}) into (\ref{lambda})  and we find:
\begin{equation}\label{eq:rLa}
  \rho_\Lambda(a)=\frac{{\rho}_I}{\alpha}\,\frac{1}{\left[1+D\,a^{4}\right]^{2}}\,.
\end{equation}
Then solving (\ref{frie33}) we obtain:
\begin{equation}\label{eq:rhor}
 \rho_r(a)=\frac{{\rho}_I}{\alpha}\,\frac{D\,a^{4}}{\left[1+D\,a^{4}\right]^{2}}\,.
\end{equation}
Here $\rho_I=3H_I^2/\kappa^2$ is the critical density in the inflationary epoch. As it is obvious from the above expressions,  there is no singularity in the initial state: the Universe starts at $a=0$ with a huge vacuum energy density $\rho_I/\alpha$ (and zero radiation) which is progressively converted into relativistic matter. In the asymptotic radiation regime we indeed retrieve the standard behavior $\rho_r \sim  a^{-4}$ with essentially negligible vacuum energy density: $\rho_\Lambda\sim a^{-8}\ll\rho_r$. Graceful exit is, therefore, implemented.

Subsequently the radiation component becomes subdominant and the matter
dominated era appears. This is the point when the $c_{0}/H^{2}$ term in
Eq.(\ref{HE}) surfaces and starts to dominate over $\alpha
H^{2}/H_{I}^{2}$ because the early de Sitter era is left well
behind ($H \ll H_{I}$). In this case Eq.(\ref{lambda}) boils down to
$\Lambda(H)=\tilde \Lambda_{0}+3\nu(H^{2}-H_{0}^{2})$ which
corrects the concordance $\Lambda$CDM model \emph{\`a posteriori}. Notice, that
\begin{equation}\label{l0cdm}
\tilde \Lambda_{0}=3c_{0}+3\nu H^{2}_{0}
\end{equation}
 is the vacuum (cosmological
constant) energy density at the present time, which is positive, and should not be confused with the negative tree-level $\Lambda_0$ of the SUGRA model (\ref{l0}). This can be understood by studying the evolution of the universe at a time after recombination,
therefore consisting of dust ($\omega_m=0$) plus the running vacuum
fluid with $H\ll H_I$. In this case,  using $d/dt=aH\,d/da$,
we can rewrite Eq.(\ref{HE}) as
$$
a\,\frac{dH^2}{da}+3(1-\nu)H^2-3\,c_0=0\,,
$$
The solution satisfying the boundary condition $H=H_0$ at present
($a=1$) is:
$$
H^{2}(a) = \frac{H_0^2}{1-\nu}
\left[(1-\Omega_{\Lambda}^{0})\,a^{-3(1-\nu)}+\Omega_{\Lambda}^0-\nu
\right]\,.
$$
Note that the aforementioned boundary condition fixes the value of
the parameter $c_0$ as follows: $c_0=H_0^2(\Omega_{\Lambda}^0-\nu)$.
For $\nu=0$ we correctly recover the behavior of the $\CC$CDM. However, for small $\nu$ the Universe possesses a mildly evolving vacuum energy that could appear as dynamical dark energy without invoking spurious scalar fields.
Furthermore, the above vacuum model is in agreement with the latest cosmological data and
it predicts a growth rate of clustering which is in agreement with
the observations (for more details concerning the late dynamics, see
\cite{GoSolBas2015,BPS09,GoSol2015,GrandeET11}).

Let us note that the main stage of the cosmic evolution where we can match the SUGRA model of Sect. IV with the RVM is the early period comprising inflation and the incipient radiation epoch, to which it leads after graceful exit, as described in this section. Later on the microscopic description of the SUGRA model is more difficult to analyze and we adopt here the point of view that the subsequent effective behavior of the Universe still follows the RVM flow dictated by the general RG equation (\ref{runningrho}), which, to order $H^4$, entails the dynamical vacuum energy density (\ref{lambda}). As previously mentioned, at low energies this implies that only the dynamical part $\sim H^2$ is active and may lead to interesting phenomenological implications for the dynamical DE of the current universe\,\cite{GoSolBas2015,GoSol2015,GrandeET11})

\subsection{Geometrical description: RVM versus Starobinsky}

Finally, let us focus now on some aspects of the inflationary era that are especially relevant for the present study. As we have
already mentioned, in this epoch we have a de Sitter solution  $H^2\simeq (1-\nu) H_I^2/{\alpha}=$const. Now, as previously indicated, $\dot{H}\simeq 0$ in this period and hence
$R\simeq 12H^{2}$. Finally, neglecting the matter component from the action
(\ref{action1}), which is justified in the inflationary period, and using $\Lambda(H)\simeq 3\alpha
H^{4}/H^{2}_{I}$ [see Eq.(\ref{lambda})]  we
schematically find
\begin{align}\label{steps1}
    S_{R,\Lambda}&=\int d^4 x \sqrt{-g}\,  \left[\frac{1}{2\kappa^{2}} R -\rL(H)\right] \nonumber \\
    &\sim \frac{1}{2\kappa^{2}}\int d^4 x \sqrt{-g}\,  \left(R-6\alpha\frac{H^{4}}{H^{2}_{I}} \right)~.
\end{align}
This demonstrates our point that an inflationary vacuum can be connected smoothly, under the RVM, with a late epoch $\CC$CDM Universe.
However, there is \emph{no unique} way by means of which we can associate the inflationary era RVM effective action (\ref{steps1}) to a microscopic model, which, as already mentioned, is to be expected due to the generic features of the RVM that describe classes of models and therefore
may correspond to more than one microscopic theories, as far as the exit from inflationary phase is concerned.

An interesting point concerns Eq.~(\ref{steps1}) if one replaces $H^4$ by the square of the Ricci scalar. In this case one may write
$S_{R,\Lambda} \simeq \int d^4 x \sqrt{-g}\,  \frac{1}{2\kappa^{2}} \Big(R - \alpha\frac{R^{2}}{24\, H^{2}_{I}}\Big) $.
Notice that, since $\alpha > 0$ in our case, the RVM model is \emph{not formally and directly equivalent} to a Starobinski-type model, for which the effective Lagrangian has the form (\ref{staroaction}) corresponding to a negative $\alpha$ coefficient in (\ref{steps1}).
This point has also been discussed in \cite{SolGo2015a}. The root of the problem lies in the fact that the metric tensors of the two models, (\ref{staroaction}) and (\ref{steps1}) are different, related by a non-trivial  conformal transformation (\ref{confmetric}) involving the linearising Hubbard-Stratonovich field $\varphi$, which plays the r\^ole of the ``physical'' inflaton. The RVM metric is identified with the  Einstein-frame metric $g_{\mu\nu}^E$ in (\ref{confmetric}), while the original one-loop effective SUGRA action is described in terms of the $g_{\mu\nu}$ metric.
Nevertheless, contact with Starobinsky-type models, like the one induced within the context of SUGRA model examined here, can be achieved by observing that it is precisely the passage from the Einstein to Jordan-frame actions, via (\ref{confmetric}),  which guarantees the opposite sign, relative to the Ricci scalar term, of the effective potential (\ref{staropotent}) of the Hubbard-Strstonovich inflaton field  in (\ref{steps}).
Upon making the identification for large \phantom{} $\kappa\varphi \gg 1$
\begin{equation}\label{bound}
3\, \alpha \, \frac{H^4}{\ka ^2 \, H_I^2} =  V_{\rm eff}(\varphi ) = \frac{3{\cal M}^2\left( 1 - e^{-\sqrt{\frac{2}{3}} \, \ka\varphi } \right)^2}{4\,\ka^2} \,
\end{equation}
where in the SUGRA model the scalaron mass scale ${\mathcal M}$ is given by (\ref{staroours}), one obtains the connection of the RVM with the microscopic Starobinsky-type inflationary SUGRA model.  The exit from the inflationary phase, then, which in fig.~\ref{fig:potstar} corresponds to the region of small
$\ka \, \varphi < 4$, which in the context of the SUGRA model would require detailed knowledge of the matter content of the theory, is then ``effectively'' described by the RVM evolution with the initial condition (\ref{bound}), that ``fills up'' the missing details in the exit-phase of the evolution in a rather generic manner.
{Below we compare the two cosmological models at the dynamical level.

\subsection{Scalar field description: RVM versus Starobinsky}
Although the fundamental origin of the RVM has
a root in the general structure of the effective action of
QFT in curved space-time, we cannot provide the latter
at this point, \phantom{}see \cite{ShapSol09} for an explanation. However, we can resemble
it via an effective scalar field $\phi$ using a field theoretical
language. We may call this scalar field $\phi$ as {\em vacuumon}.
Based on Friedmann's Eqs.(\ref{friedr})-(\ref{friedr2}) and
following standard lines, namely
$\rho_{\rm tot}\equiv \rho_{\phi}={\dot \phi}^{2}/2+U(\phi)$
and $p_{\rm tot}\equiv p_{\phi}={\dot \phi}^{2}/2-U(\phi)$ we arrive
at
\begin{equation}
\dot{\phi}^{2} =-\frac{2}{\kappa^{2}}\dot{H} \;, \label{ff3}
\end{equation}
\begin{equation}
\label{Vz} U=\frac{3H^{2}}{\kappa^{2}}\left(
1+\frac{\dot{H}}{3H^{2}}\right)= \frac{3H^{2}}{\kappa^{2}}\left(
1+\frac{aH^{'}}{3H}\right) \;,
\end{equation}
where $U(\phi)$ is the effective potential,
$\dot{H}=aHH^{'}$ and prime here denotes derivative with
respect to the scale factor. Integrating Eq.(\ref{ff3}) we have
\begin{equation}
\label{ppz} \phi=\int \left( -\frac{2\dot{H}}{\kappa^{2}}\right)^{1/2}
dt = \frac{\sqrt{2}}{\kappa}\int
\left(-\frac{H^{'}}{aH}\right)^{1/2}da\;.
\end{equation}

Now, for $\omega_m=1/3$ the Hubble parameter (\ref{HS1}) takes the form
\begin{equation}\label{HS11}
 H(a)=\left(\frac{1}{\alpha}\right)^{1/2}\,
\frac{H_{I}}{\sqrt{D\,a^{4}+1}}\,.
\end{equation}
Notice that, 
we have set $\nu=0$ in Eq.(\ref{HS1}), which is not important
for the study of the early universe.
Inserting Eq.(\ref{HS11}) into Eq.(\ref{ppz})
and performing the integration in the
interval $[0,a]$ we find

\begin{figure}
\includegraphics[width=0.45\textwidth]{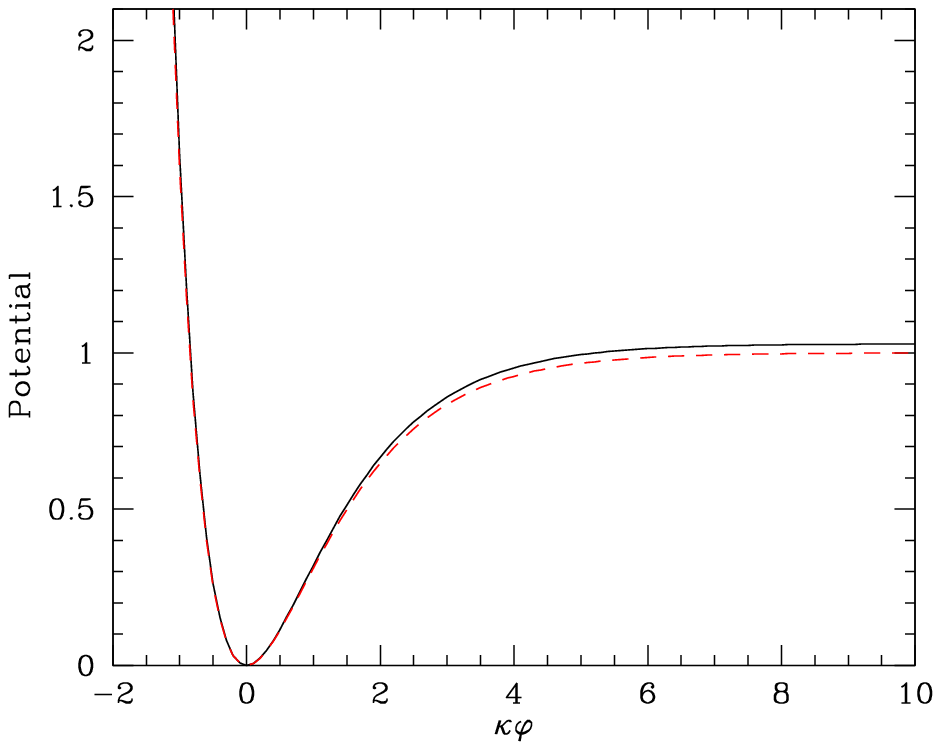}
\caption{The RVM effective potential
$\alpha \kappa^{2} U/H^{2}_{I}$ (solid line) versus
the scalaron field $\kappa \varphi$. In order to produce the curve we utilize
${\cal M} \sim M_{X} \sim 10^{16}$Gev, $H_{I}\sim 0.81\times 10^{14}$Gev
[see Eq.(\ref{upperHbic})] and $\alpha \sim 10^{-4}$.
The dashed line corresponds to the Starobinsky
effective potential (see Fig.~\ref{fig:potstar}).}
\label{fig:potstar2}
\end{figure}

\begin{eqnarray}\label{ppz1}\nonumber
\phi(a) & = & \frac{1}{\kappa}\;{\rm sinh}^{-1}\left( \sqrt{D}a^{2}\right),\\
& = & \frac{1}{\kappa}{\ln}\left(
\sqrt{D}a^{2}+\sqrt{Da^{4}+1}\right)\;.
\end{eqnarray}
In this context, utilizing Eqs.(\ref{Vz})-(\ref{HS11})
the effective potential is given by
\begin{equation}
U(a)=\frac{H^{2}_{I}}{\alpha \kappa^{2}}\;\frac{3+Da^{4}}{(1+Da^{4})^{2}},
\end{equation}
which implies
\begin{equation}
U(\phi)=\frac{H^{2}_{I}}{\alpha \kappa^{2}}\; \frac{3+{\rm
sinh}^{2}(\kappa \phi)} {[1+{\rm sinh}^{2}(\kappa \phi)]^{2}}
\end{equation}
or
\begin{equation}
\label{Pott}
U(\phi)=\frac{H^{2}_{I}}{\alpha \kappa^{2}}\; \frac{2+{\rm
cosh}^{2}(\kappa \phi)} {{\rm cosh}^{4}(\kappa \phi)}  \;.
\end{equation}
At this point we would like to pose the following dynamical
question: {\em Under what conditions
the RVM potential is equal to that of Starobinsky, namely
$U(\phi)=V_{\rm eff}(\varphi)$ in Eq. (\ref{bound})?}

Equating  the right-hand-side of
Eqs.(\ref{staropotent}) and (\ref{Pott}), after some calculations, we can express
the {\em vacuumon} field in terms of the scalaron
\begin{equation}\label{XXXa}
\phi(\varphi)=\frac{1}{\kappa}{\rm ln}\left[ \chi(\varphi)+\sqrt{\chi(\varphi)^{2}-1}\right]
\end{equation}
where $\chi(\varphi)>1$ as easily shown  (hence no restrictions on the scalar fields) and it is given by
\begin{equation}
\label{XXX}
\chi(\varphi)=\left[ \frac{1+\sqrt{1+8F(\varphi)}}{2F(\varphi)}\right]^{1/2}~,
\end{equation}
$$F(\varphi)=\frac{3\alpha {\cal M}^2\left( 1 - e^{-\sqrt{\frac{2}{3}} \, \ka\varphi } \right)^2}{4\,H^{2}_{I}} = \frac{\alpha \kappa^{2}}{H^{2}_{I}}V_{\rm eff}(\varphi)>0 \;.$$

In Fig.~\ref{fig:potstar2} we present the RVM effective potential $\alpha \kappa^{2} U/H^{2}_{I}$
(solid curve) as a function of $\kappa \varphi$. In the same figure we plot
the effective Starobinsky (dashed curve) potential which is showed in Fig.~\ref{fig:potstar}.
From the comparison it becomes clear that,
although the RVM and Starobinsky models
live in different geometrical backgrounds, namely GR and
$R^{2}$, the two models are {\em similar} from the point of
view of those features of inflation that can be described by an
effective scalar-field dynamics. However, in other important aspects
they are different. We should mention that the RVM model provides a simple description of the graceful exit and reheating problem, see
\,\cite{LBS2013,BLS2013,SolaGRF2015,LBS2016} for details. As for the Starobinsky model, the reheating of the universe after the exit of the inflationary phase it has been discussed for example in \cite{ReheatStaro}. In contrast to the
Starobinsky model, a general effective action from where the RVM can be
derived is not known\,\cite{ShapSol09}, and currently this has been achieved only in some cases~\cite{Fossil07}.}

\section{Conclusions \label{sec:5}}

In the light of the latest Planck+Bicep2
results \cite{Bicep2-Planck2015} it has been
proposed that the Starobinsky inflation plays a key role because it
fits quite well the Cosmic Microwave Background (CMB) data on
inflation. In the present paper we have further investigated the class of the running vacuum models (RVM) \cite{JSPRev2013}  (based on renormalization-group approach in curved spacetimes)  and their implications on the inflationary universe\,\cite{LBS2013,BLS2013}. In particular, we have addressed the possibility that they can mimic both the original Starobinsky model and the spontaneously broken SUGRA models
based on dynamically induced gravitino condensates\,\cite{emdyno}.

We have shown that the vacuum energy density  $\rL (H)$ of these SUGRA models can be expressed as an even power series (\ref{runningrho}) of the Hubble
parameter, which can be naturally truncated at the $H^4$ term.  This is exactly the generic form expected in the simplest class of running vacuum models and therefore we can apply the known implications of these models for inflation\,\cite{LBS2013,BLS2013}. Namely, after computing the modified form of the Friedmann equation, we find that the physics of inflation
(which in our case occurs for $H\simeq H_I$, a value associated to the spontaneously broken SUGRA model) is mainly described by the $H^4$-term. Furthermore, being  $H^4$ of order  $R^{2}$, we can trace some relationship of this model with Starobinsky inflation, although of course there is not a full identification or equivalence. Most noticeably we point out the distinguishing feature that within  the entire class of running vacuum models -- and hence, in particular, the SUGRA model that we have studied (which adapts to the same pattern) --
the RVM performs successful graceful exit from the inflationary phase into the standard radiation regime\,\cite{LBS2013,BLS2013}. This feature is characteristic of the running type of vacuum models, in contrast to the original Starobinsky model. {Nevertheless, we have also shown here that the RVM model admits a scalar field description as well, via the vacuumon field, and its potential can be made equivalent to the Starobinsky potential upon appropriate scalar field redefinitions, despite the fact that the geometric backgrounds of the two models are very different. This dynamical equivalence implies that the two models \phantom{} should provide the same
inflationary features, at least in all the aspects that can be described through an effective scalar field potential.
Not so in other aspects which may differ from one model to the other. \phantom{} In particular let us emphasize
that the Starobinsky model derives from a local effective action whereas the structure of
the effective action in the general case RVM is not presently known, except in particular cases in which it is found to be non-local.

The low energy physics, on the other hand, and in particular the evolution of the Universe in the current epoch,  is determined by the constant additive term of $\rL(H)$ and the power $H^2$, which provides a remnant dynamical evolution still in our days, which is of the form $\rL(H)=\rho_{\Lambda}^0+(3\,\nu/\kappa^2)(H^2-H_0^2)$. Such evolution is mild because the coefficient  of $H^2$ is small (it is interpreted as the $\beta$-function coefficient of the running vacuum energy at present). The signature of the RVM at present is precisely that mild quadratic dynamical behavior of the vacuum energy density around the current value $\rho_{\Lambda}^0$ which is parameterized by the small parameter $\nu$. The model has  been thoroughly put to the test recently and it allows values of $|\nu|={\cal O}(10^{-3})$\,\cite{GoSolBas2015,BPS09,GoSol2015,GrandeET11}). On the other hand, its successful performance in describing the physics of the early Universe (in particular the graceful exit of the inflationary phase into the standard radiation one) is also quite encouraging, especially after realizing that specific QFT models lead to this kind of behavior. In this paper we have shown that SUGRA models with a dynamically induced massive gravitino phase lead to the RVM behavior and therefore provide a strong support for a fundamental description of the cosmic history.

Finally, we would like to stress that, in the context of the running vacuum model, the universe evolution, and especially its
accelerated phase either during inflation or at late times,
is not attributed to an \textit{ad hoc} scalar field,
or to a modification of the gravitational interaction, but rather
arises from the
modification of the vacuum itself, which is endowed with a dynamical nature. Remarkably, the SUGRA framework studied here provides a concrete realization of this possibility within the fundamental context of quantum field theory in curved spacetime.

\vspace{0.5cm} \textbf{Acknowledgments.} SB acknowledges support by
the Research Center for Astronomy of the Academy of Athens in the
context of the program  ``{\it Tracing the Cosmic Acceleration}''.
The work  of N.E.M. is supported in part by the London Centre for
Terauniverse Studies (LCTS), using funding from the European
Research Council via the Advanced Investigator Grant 267352 and by
STFC (UK) under the research grant  ST/L000326/1. The work of JS has
been partially supported by FPA2013-46570 (MICINN), Consolider grant
CSD2007-00042 (CPAN), 2014-SGR-104 (Generalitat de Catalunya) and MDM-2014-0369 (ICCUB).


\end{document}